\documentclass[chaos,amsmath,amssymb,showpacs,groupaddress,twocolumn]{revtex4}
\usepackage{graphicx}
\usepackage{bm}
\begin{document}
\title{Topological field theory of dynamical systems}
\author{Igor V. Ovchinnikov}
\email{iovchinnikov@ucla.edu, igor.vlad.ovchinnikov@gmail.com}
\affiliation{Department of Electrical Engineering, University of California at Los Angeles, Los Angeles, CA, 90095-1594}
\begin{abstract}
Here, it is shown that the path-integral representation of any stochastic or deterministic continuous-time dynamical model is a cohomological or Witten-type topological field theory, \emph{i.e.}, a model with global topological supersymmetry ($\mathcal Q$-symmetry). As many other supersymmetries, $\mathcal Q$-symmetry must be perturbatively stable due to what is generically known as non-renormalization theorems. As a result, all (equilibrium) dynamical models are divided into three major categories: Markovian models with unbroken $\mathcal Q$-symmetry, chaotic models with $\mathcal Q$-symmetry spontaneously broken on the mean-field level by, \emph{e.g.}, fractal invariant sets (\emph{e.g.}, strange attractors), and intermittent or self-organized critical (SOC) models with $\mathcal Q$-symmetry dynamically broken by the condensation of instanton-antiinstanton configurations (earthquakes, avalanches \emph{etc}.) SOC is a full-dimensional phase separating chaos and Markovian dynamics. In the deterministic limit, however, antiinstantons disappear and SOC collapses into the "edge of chaos".  Goldstone theorem stands behind spatio-temporal self-similarity of $\mathcal Q$-broken phases known under such names as algebraic statistics of avalanches, 1/f noise, sensitivity to initial conditions \emph{etc}. Other fundamental differences of $\mathcal Q$-broken phases is that they can be effectively viewed as quantum dynamics and that they must also have time-reversal symmetry spontaneously broken. $\mathcal Q$-symmetry breaking in non-equilibrium situations (quenches, Barkhausen effect, \emph{etc}) is also briefly discussed. \end{abstract}
\maketitle
{\bf Many fundamental aspects of the theory of nonlinear dynamical systems have already been established. Nevertheless, our understanding of nonlinear dynamics may benefit from turning to alternative approaches. Here, it is shown that one of such alternatives is Witten-type or cohomological topological field theories (W-TFTs). \cite{TFT, WTFT,Affleck,WittenForms, Anselmi,Labastida,DynamicalBreakingOfSUSY,
InstantonBRSTBreaking,BRSTSymmetryBroken,Frenkel} It turns out that any dynamical model in its path-integral representation is a W-TFT. This approach allows to clarify the physical essence of Intermittency also known as self-organized criticality (SOC, for a review see, \emph{e.g.}, Refs.[\onlinecite{SOC}]) and establish its connections with the other two fundamental concepts of Chaos and Markovianity.}

\section{Introduction}
For the last two decades, mathematical physicists and mathematicians have formulated and have been developing mathematical constructions known as topological field theories (TFTs, for a review see, e.g., Ref.[\onlinecite{TFT}]). TFTs come in two types: Schwartz-type or quantum TFTs and Witten-type or cohomological TFTs (W-TFTs). Quantum TFTs already found their applications - they are believed to stand behind exotic low-temperature phases of condensed matter systems such as fractional quantum Hall effects and superconductors (see, e.g., Ref.[\onlinecite{STFT}] and Refs. therein). As to the W-TFTs, to the best of our knowledge they are still used only for purely mathematical purposes. From the discussion in this paper it will follow that W-TFTs is actually the path-integral version of the dynamical systems theory. The later, in turn, has many applications in modern science.

It is well known that the most general Parisi-Sourlas-Wu (PSW) stochastic quantization procedure \cite{ParisiSourlas,ParisiSourlasWu} applied to Langevin equations leads to N=2 (quasi-)Hermitian supersymmetric models (Witten models, see, \emph{e.g.}, Refs. [\onlinecite{TFT}, \onlinecite{Zinn-Justin}]). It is also known that deterministic conservative (classical, Hamilton) dynamical systems viewed as path-integrals also possess supersymmetry. \cite{Gozzi,Kurchan} The above supersymmetries are of topological origin. \cite{Labastida} In this paper, it is demonstrated that topological supersymmetry is pertinent to any deterministic or stochastic continuous-time dynamical model in its path-integral representation. Based on this and on the possibility of the spontaneous breakdown of the topological supersymmetry by two different mechanisms, a generic phase diagram for dynamical models is proposed. The phase diagram is given in Fig.\ref{Figure4}b.

As quantum field theories, W-TFTs that show up on PSW quantization are non-Hermitian. Therefore, our proposition relies strongly on recent developments in the theory of non-Hermitian quantum dynamics. \cite{Bender,Mostafa,SinhaRoy}

The paper is organized as follows. In Sec.\ref{StochQuantization}, it is demonstrated that any stochastic or deterministic continuous dynamical model is a W-TFT. In Sec.\ref{WhiteNoise}, we focus on models with white noise. Non-Hermitianity of the models (Sec.\ref{NonHermitianity}), their spectrum and physical states (Sec.\ref{PathIntegralSection}), \emph{etc.}, are addressed. In Sec.\ref{VacuaDetermin}, the meaning of the ground states in deterministic limit is analyzed. It is shown that in cases of unbroken topological symmetry the collection of (bra's) ket's of the perturbative ground states is the representation of the (anti-)instantonic CW-complex of the phase space. In turn, the global ground states represent stable and unstable manifolds that intersect on invariant manifolds. Sec. \ref{UnbrokenSymmetry} is devoted to models with unbroken topological symmetry, and, in particular, to Langevin models. It is also discussed how W-TFT's with no physical quantum excitations represent stochastically fluctuating dynamical models (Sec.\ref{Explanation}). In Sec.\ref{SpontanBroken}, models with spontaneously broken topological symmetry are addressed. Qualitative difference of dynamics with spontaneously broken topological symmetry is discussed in general terms in Sec.\ref{QualitativeDifference}. Particularly, it is proposed that such concepts of dynamical systems theory and complexity theory as self-similarity, sensitivity to initial conditions, non-Markovianity \emph{etc}. are actually mere consequences of the Goldstone theorem. Two mechanisms of spontaneous topological symmetry breaking are identified: on the mean-field level, \emph{e.g.}, in the deterministic limit, and due to the dynamical condensation of instantons and antiinstantons. It is argued that these two mechanisms corresponds respectively to chaos (Sec.\ref{Chaos}) and intermittency/SOC (Sec.\ref{SOC}). In Sec. \ref{NonEquilibrium}, non-equilibrium dynamics such as quenches or Barkhausen effect is briefly addressed in the context of W-TFTs. Sec. \ref{Conclusion} concludes the paper.

\section{Stochastic quantization}
\label{StochQuantization}
Dynamical systems are those that are defined by specifying their equations of motion. This is the most general class of models as compared to, say, the equations of motions that follow from least action principle.

The equations of motion can be either step-like equations or time-continuous equations. This paper deals with time-continuous evolutions only. Continuous dynamics can be referred to as physical dynamics since time is always continuous in physical systems.

The equations of motion can be either stochastic or deterministic (partial) differential equations (SDE or DDE). DDE's can be looked upon as SDE's with zero-variance noises. Therefore, it suffices to construct a path-integral formulation for an SDE, and the theory of a corresponding deterministic model will follow by sending the parameters of the noise to their deterministic limit. Thus, SDEs are of primary interest.

The procedure of building a partition function out of an SDE is called stochastic quantization. There are two major stochastic quantization procedures: PSW method \cite{ParisiSourlas,ParisiSourlasWu} and Martin-Siggia-Rose (MSR) method. \cite{MSR} As is discussed below, MSR is an approximation to the most general PSW method that we use in this paper.

Consider an SDE in its most general form:
\begin{eqnarray}
F^i(R,\varphi) = \xi^i(R,\varphi).\label{SDEGeneral}
\end{eqnarray}
Here $R\in B$ with $B$ being some Riemannian base space having the meaning of spacetime and $\varphi\in M(\Sigma, B)$, with $M$ being an infinite-dimensional space of all maps from $B$ to a topological target manifold, $\Sigma$. In Eq. (\ref{SDEGeneral}), $F$'s have the meaning of a DDE and $\xi$'s is the stochastic noise. $F$'s are some functionals of $\varphi$'s and may have explicit dependence on the base. Both $F$'s and $\xi$'s belong to the tangent space of $M$, \emph{i.e.},  $F^i(R,\varphi)\delta/\delta\varphi^i(R), \xi^i(R,\varphi)\delta/\delta\varphi^i(R)\in T_{\varphi^i(R)} M$. $F^i$'s and $\xi$'s can also be thought of as sections of the tangent bundle $TM$.

The noise must obey an important physical condition. It must experience no feedback from the system. If there is such a feedback, the noise should be viewed as a part of the system rather than as an external source of stochasticity. This condition, however, does not necessarily suggest that the stochastic correlators of the noise have no functional dependence on $\varphi$. For example, below we concentrate on models where the noise is decoupled from the system but its correlators are functions of $\varphi$. This possible dependence of the correlators of the noise on $\varphi$'s is emphasized by the explicit dependence of the noise on $\varphi$ in the rhs of Eq.(\ref{SDEGeneral}).

The above condition only suggests that there exists an invertible transformation of $TM\to TM$:
\begin{eqnarray}
(\varphi,\xi) \to (\tilde\varphi,\tilde \xi),\label{XiTransform}
\end{eqnarray}
that decouples the stochastic variables from $\varphi$'s. An example of such a transformation is given in Eq.(\ref{TransfromationExmple}). The partition function of new stochastic variables is:
\begin{eqnarray}
\langle\langle 1 \rangle\rangle = \iint_{[\tilde\xi]} 1 \cdot e^{-S_\text{noise}(\tilde\xi)},\label{PartitionFunctionNoise}
\end{eqnarray}
where $S_\text{noise}$ is independent of $\tilde\varphi$'s.

In Eq.(\ref{PartitionFunctionNoise}) and in the following the double integral sign denotes functional integration. $\langle\langle ... \rangle\rangle$ is a common notation denoting a vacuum expectation value in the theory of quantum non-Hermitian models, which the models we consider here will turn out to be. We accept this notation from the very beginning of our discussion.

Now, the original SDE (\ref{SDEGeneral}) can be rewritten as:
\begin{eqnarray}
\tilde F^a(R,\tilde\varphi) = \tilde \xi^a(R),\label{TransformedSDE}
\end{eqnarray}
where $\tilde F^a(R,\tilde \varphi) = \tilde \xi^a (\varphi, F(R,\varphi))$ is the functional transformation from Eq.(\ref{XiTransform}) applied to $F$'s. Eq.(\ref{transfromSDEWhite}) below is an example of the transformed SDE (\ref{TransformedSDE}).

We assume that the noise is "decent" and for any, $X_a(R)$, there exist a unique and well-defined functional
\begin{eqnarray}
S^{-1}_\text{noise}(X) = \log \langle\langle e^{\int_{R}\tilde\xi^a(R)X_a(R)} \rangle\rangle\nonumber\\ = \sum_{n=2}^\infty \int_{R_1...R_n}C_{(n)}^{a_1...a_n}(R_1...R_n)\prod_{i=1}^n X_{a_i}(R_i)/n!,\label{InversePartitionFunction}
\end{eqnarray}
where $C$'s are the irreducible stochastic correlators for $\tilde\xi$'s. Here and in the following the factors representing the metric on $B$ are assumed whenever appropriate.

The inverse functional transformation in Eq. (\ref{InversePartitionFunction}) is well defined, \emph{e.g.}, when the noise has only one "vacuum", \emph{i.e.}, only one solution for $\delta S_\text{noise}/\delta \tilde\xi^a(R)=0$. This vacuum must be stable and non-degenerate so that all the eigenvalues of $\delta^2 S_\text{noise}/\delta \tilde \xi^a(R)\delta \tilde \xi^b(R')$ are positive. Such "instanton-free" noises can be called Markovian in the context of this paper.

The first step toward a W-TFT, which to the best of our knowledge belongs to Parisi and Sourlas, \cite{ParisiSourlas} is very natural. It is the realization of the fact that the only partition function a stochastic system may in principle have is that of its noise, Eq.(\ref{PartitionFunctionNoise}). Indeed, the partition function is the summation over all the realizations of a stochastic process, which is actually $\tilde \xi$. Furthermore, if we want that the "number" of the noise's degrees of freedom be the same as the number of the system's degrees of freedom we must consider only temporally periodic boundary conditions. This is nothing else but the equilibrium situation that we will consider mostly.

The second step is to rewrite the partition function in terms of $\tilde\varphi$'s rather than $\tilde\xi$'s. Let us now recall that in the theory of supersymmetric models there is a concept of Nicolai maps. \cite{TFT,Nicolai} It says that for theories with global supersymmetry one can come up with such a transformation of bosonic variables, that the partition function will fold into that of a noise because the Jacobian of the variable's transformation will cancel the fermionic determinant. What we need to do now is essentially the opposite. We have to unfold the partition function of the noise into some other model and rightfully anticipate that the model will possess a global supersymmetry - topological supersymmetry ($\mathcal Q$-symmetry).

The partition function of the noise is rewritten as
\begin{eqnarray}
\langle\langle 1 \rangle\rangle = \iint_{ [\tilde\varphi \tilde\xi] } \delta(\tilde F^a - \tilde \xi^a)
J e^{-S_\text{noise}(\tilde \xi)},\label{TransformationVar}
\end{eqnarray}
where $J = \det(\delta {\tilde F}^a(R,\tilde \varphi) / \delta \tilde \varphi^b(R'))$ is the Jacobian of the variable transformation. The $\delta$-functional limits the integration over $M$ only to the solutions of the SDE, that serves as an "inverse" Nicolai map. In order to bring Eq.(\ref{TransformationVar}) back to Eq.(\ref{PartitionFunctionNoise}), one must integrate out $\tilde\xi$'s and notice that the outcome is actually Eq.(\ref{PartitionFunctionNoise}) with $\tilde\xi$'s substituted by $\tilde\varphi$'s according to the variable transformation specified by Eq.(\ref{TransformedSDE}).

The Jacobian of the variable transformation is important. Without it, Eq.(\ref{TransformationVar}) is not the same as Eq.(\ref{PartitionFunctionNoise}) and thus does not represent the stochastic process under consideration. Neglecting the Jacobian is the mathematical essence of the MSR stochastic quantization procedure. Therefore, MSR picture can only serve as an approximation, while the PSW stochastic quantization procedure, the one we use here, is always correct.

In the standard fashion, one introduces the Lagrange multiplier, $B_a$, to incorporate the $\delta$-functional:
\begin{subequations}
\label{help}
\begin{eqnarray}
\delta(\tilde F^a - \tilde \xi^a) \sim \iint_{[B]} e^{i\int_R B_a(\tilde F^a - \tilde \xi^a)}.
\end{eqnarray}
Here and in the next formula, $\sim$ denotes equality up to an unimportant constant. One also introduces the set of the Fadeev-Popov ghosts, $\chi^a\bar\chi_b$, integration over which provides the desired Jacobian
\begin{eqnarray}
J \sim \iint_{[\chi\bar\chi]} e^{-i\int_{RR'}\bar\chi_a (\delta {\tilde F}^a/ \delta \tilde \varphi^b) \chi^b}.
\end{eqnarray}
\end{subequations}
On inserting Eqs.(\ref{help}) into Eq.(\ref{TransformationVar}) one obtains
\begin{eqnarray}
\langle\langle 1 \rangle\rangle = \iint_{[\tilde\xi \Phi]} e^{i\left\{\mathcal{Q}, \int_{R}\bar\chi_a(\tilde F^a - \tilde\xi^a)\right\} - S_\text{noise}(\tilde\xi)}.\label{IntermediatePartitionFunction}
\end{eqnarray}
Here and in the following $\Phi$ stands for the collection of all the fields $\tilde\varphi B\chi\bar\chi$. These fields constitute what could be called the supersymmetric extension of $TM$. In Eq.(\ref{IntermediatePartitionFunction}), we introduced the operator of the gauge-fixing Becchi-Rouet-Stora-Tyutin supersymmetry, which is the $\mathcal Q$-symmetry:
\begin{eqnarray}
\{\mathcal{Q},X\} = \int_{R} \left(\chi^a(R)\hat \delta_{\tilde\varphi^a(R)}+B_a(R)\hat\delta_{\bar\chi_a(R)}\right)X.\label{Qsymmetry}
\end{eqnarray}
Here and in the following $\hat \delta$ (or $\hat \partial$) denotes functional (or partial) differentiation over its subscript. Operator $\mathcal Q$ is a bi-graded differentiation, $\{\mathcal {Q}, XY\} = \{\mathcal {Q}, X\} Y + (-1)^{m} X \{\mathcal {Q},Y\}$, where $m$ is the ghost degree of $X$. $\mathcal Q$ is nilpotent, $\{\mathcal {Q},\{\mathcal {Q},X \}\}=0$, for any $X$.

Out integration of the noise in Eq.(\ref{IntermediatePartitionFunction}) leads to
\begin{eqnarray}
\langle\langle 1 \rangle\rangle = \iint_{[\Phi]} e^{i\left\{\mathcal{Q},\int_{R}\bar\chi_a \tilde F^a\right\} + S^{-1}_\text{noise} (-i \{\mathcal{Q},\bar\chi_a\})}.\label{IntermediatePartitionFunction1}
\end{eqnarray}
where the last term in the action is defined in Eq.(\ref{InversePartitionFunction}). This term is a functional of only $\mathcal Q$-exact pieces (of the form $\{\mathcal{Q}, X \}$) and is $\mathcal Q$-exact itself because of the nilpotency of $\mathcal Q$: $\{{\mathcal Q}, X\}\{{\mathcal Q}, Y\}...=\{{\mathcal Q}, X\{{\mathcal Q}, Y\}...\}$. Thus,
\begin{eqnarray}
\langle\langle 1 \rangle\rangle = \iint_{[\Phi]} e^{i\{ \mathcal{Q}, \Theta \}},\label{WTFTPartitionFunction}
\end{eqnarray}
with the so-called gauge fermion given by
\begin{eqnarray}
\Theta &=& \int_R \bar\chi_a(R)\tilde F^a(R) +
\sum_{n=2}^\infty (-i)^{n+1}\int_{R_1...R_n}\bar\chi_{a_1}(R_1)\nonumber\\
&&\times C_{(n)}^{a_1...a_n}(R_1...R_n) \prod_{i=2}^n B_{a_i}(R_i)/n!.
\label{GaugeFermion}
\end{eqnarray}
The last term in this expression comes from the noise and for deterministic models it vanishes.

A $Q$-exact action is the unique feature of W-TFTs that all look like a gauge fixing of "nothing". That the PSW method leads to a W-TFT comes with no surprise. The point is that we are computing the partition function of the "ignorant" external noise, which does not really care about the lhs of Eq.(\ref{TransformedSDE}). Therefore, the deformation of the SDE must not result in any changes in the partition function - a feature pertinent to W-TFTs.

Topological nature of models with $\mathcal Q$-exact actions and with non-trivial topological content can be in particular revealed by the provided possibility to calculate on instantons certain topological invariants as expectation values of collections of certain $\mathcal Q$-closed operators also known as BPS (Bogomol'nyi-Prasad-Sommerfield) operators. \cite{Frenkel} Nevertheless, even W-TFTs with topologically trivial content are of topological origin - in Ref.\cite{Labastida}, any model with a $\mathcal Q$-exact action has been given an interpretation of a generalized Morse theory on $M$. Some new mathematical insights on W-TFTs have been recently developed in Ref.[\onlinecite{Frenkel}].

\begin{figure}[t] \includegraphics[width=8.0cm,height=3cm]{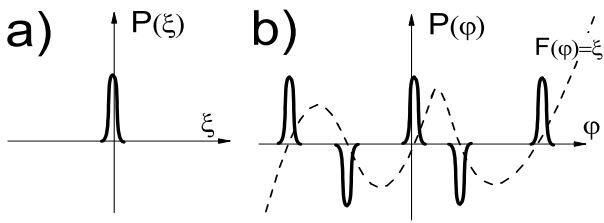}
\caption{\label{Figure1}
{\bf (a)} Graphical explanation of how negative probabilities show up in stochastic quantization. Original noise has a positive delta-function-like probability distribution. {\bf (b)} Corresponding to a highly nonlinear many-to-one variable transformation, $F(\varphi)=\xi$ (dashed curve), that has the meaning of an SDE, the probability distribution, $P(\varphi)$ (solid think curve), is negative in those regions where the Jacobian $\partial F /\partial \varphi <0$.}
\end{figure}

\subsection{Connection to Negative Probabilities}
\label{NegativeProb}

Stochastic quantization seems to have a close connection with the concept of negative probabilities (see, \emph{e.g.}, Ref.[\onlinecite{NegativeProb}] and Refs. therein). This concept shows up naturally on the PSW quantization.

Consider, \emph{e.g.}, a nearly-deterministic Gaussian noise, $\xi\in\mathbb{R}^1$, in a (0+0) theory. The partition function is:
\begin{eqnarray}
\langle\langle 1 \rangle\rangle = \int_\xi P(\xi) = \int_\varphi P(\varphi) = i \int_{\Phi}e^{i\{ \mathcal {Q},\Theta\}}.\label{Example}
\end{eqnarray}
Here, $P(\xi)\sim e^{-\xi^2/2\epsilon}$ with $\epsilon\to 0$, the second integral is over a new variable that is related to $\xi$ through some many-to-one function, $\xi=F(\varphi)$, $P(\varphi) = P(F(\varphi)) J(\varphi)$ is the new probability density with $J(\varphi)=\partial F(\varphi)/\partial\varphi$ being the Jacobian of the variable transformation, and the last equality sign establishes the connection to the discussion in the previous subsection with $\Theta = \bar\chi(F(\varphi)+i\epsilon B/2)$.

New probability distribution is given in Fig.(\ref{Figure1}). In those regions where the Jacobian is negative, $P(\varphi)$ is also negative. These negative probabilities are necessary for the partition function to be equal to the partition function of the original stochastic process. They always appear paired with new positive probabilities and thus are clearly of topological origin - the simplest realization of the Poic\'are-Hoft theorem.

A mathematician would probably say that $P(\varphi)$ is a pullback of $P(\xi)$ by the irreversible many-to-one map, $F(\varphi):\varphi\to\xi$. The original $P(\xi)$ is non-negative and thus can be interpreted as a volume form for $\xi$. From this point of view, the stochastic quantization is the procedure of "borrowing" the volume from the noise, so that the total volume in the new variables (the total probability to exist) remains the same as in Eq.(\ref{Example}).

This simple example suggests the following stochastic interpretation of negative probabilities. If a stochastic variable has negative probabilities, it means that its partition function actually represents/calculates a partition function of yet another "physical" stochastic variable (noise) with ordinary positive probability. The two stochastic variables must be related through some irreversible many-to-one map, \emph{e.g.}, through a highly nonlinear SDE.

On the side of stochastic dynamics, our temporary understanding of the negative probabilities is this. For a specific configuration of the noise, a nonlinear SDE may have a multitude of solutions that in the high energy Physics terms are called Gribov copies. The system has a freedom to chose which one of the Gribov copies is going to be realized. We can not know which one of the Gribov copies the stochastic system is going to chose. Only this chosen copy must contribute to the partition function. At the same time, we have to integrate over all the possible configurations of the systems variables (over the entire $M$) and all the Gribov copies are going to contribute. This would necessary lead to overcounting unless some of the Gribov copies are contributing -1 instead of 1 due to that the probability density is negative in the corresponding regions of $M$. Thus, negative probabilities represent the freedom of a stochastic system to chose among various competing solutions of its SDE. From the forthcoming discussion, it follows that this freedom gets physically realized only in $\mathcal Q$-broken phases.

\section{White noise case}
\label{WhiteNoise}
For the following discussion we do not need the generality of the previous section. We assume that spacetime, $R = (t,x) \in B=T\times S$, where $T$ is time and the space, $S$, is flat (\emph{e.g.}, a torus $S=T^d$, \emph{i.e.}, periodic boundary conditions). In equilibrium situations $T=\mathbb{S}^1$, while in non-equilibrium cases $T=\mathbb{R}^1$. We discuss mostly the equilibrium situations.

SDE has only the simplest temporal non-locality:
\begin{eqnarray}
\partial_t\varphi^i + A^i = \xi^i,\label{reducedSDE}
\end{eqnarray}
where the flow, $A^i$, is some functional of $\varphi$ at this specific moment of time. $A$ can be though of as a vector field over the (infinite-dimensional) phase space, $H$, of all maps from the space to the target: $H=\{\varphi^i(x):\mathcal{S}\to\Sigma\}$.

The noise is assumed Gaussian and white:
\begin{eqnarray}
\langle \langle \xi^i(R) \xi^j(R')\rangle \rangle &=& \delta(R-R')g^{ij}(\varphi(R)),
\end{eqnarray}
where $g^{ij}$ has the meaning of the metric on $\Sigma$, which is a function of only $\varphi^i$'s at this specific $R$.

Eq.(\ref{reducedSDE}) and/or Eq.(\ref{WTFTPartitionFunction}) with Eq.(\ref{GaugeFermionReduced}) below represent a very wide class of stochastic dynamical models. In particular, this class includes systems defined by SDE's that are not first-order in time-derivative. One can bring such higher-order SDE's to the form of Eq. (\ref{reducedSDE}) by the introduction of new fields. In case of Kramer' s equation this works like this: $\partial_t^2\varphi - A = \xi$ is rewritten as $\partial_t p - A = \xi, \partial_t \varphi - p = \xi'$, where $\xi'$ is a zero-variance Gaussian noise.

The transformation that decouples the noise from $\varphi$'s has the following form
\begin{eqnarray}
\tilde \xi^a(R) = e^a_i(\varphi(R)) \xi^i(R),\label{TransfromationExmple}
\end{eqnarray}
where $e$'s have the meaning of vielbeins: $e^a_i \delta_{ab}e^b_j  =g_{ij}$ and $e^i_a \delta^{ab}e^j_b =g^{ij}$. The transformed SDE from Eq. (\ref{TransformedSDE}) takes the following form:
\begin{eqnarray}
e_i^a(\partial_t\varphi^i+A^{i}) = \tilde\xi^a.\label{transfromSDEWhite}
\end{eqnarray}
Straightforward application of the PSW stochastic quantization procedure from the previous section to the partition function of the white noise:
\begin{eqnarray}
\langle\langle 1 \rangle\rangle = \iint_{[\tilde\xi]} e^{-\int_R (\tilde\xi^a)^2/2},
\end{eqnarray}
leads to a W-TFT for the collection of fields, $\Phi = (\varphi^i, \chi^i, B_a, \bar\chi_a)$, with a $\mathcal Q$-exact action, $S = \{\mathcal{Q}, \Theta\}$, defined by the gauge fermion, $\Theta = \int_R \bar\chi_a(e^a_i(\partial_t \varphi^i + A^i) +i \delta^{ab}B_b/2)$, and the supersymmetry operator $\{\mathcal Q,X\} = \int_{R}(\chi^i(R) \delta_{\varphi^i(R)} + B_a(R) \hat\delta_{\bar\chi_a(R)})X$.

It is more convenient, however, to introduce a new set of fields, $\Phi = (\varphi^i, \chi^i, B_i, \bar\chi_i)$, where $\varphi^i$ and $\chi^i$ are the same, while $\bar\chi_i = \bar\chi_a e^a_i$ and $B_i = \{\mathcal{Q},\bar\chi_i\} = B_a e^a_i - \bar\chi_k \Gamma^k_{il} \chi^l$ with $\Gamma^k_{il} = e^k_a (e^a_i)'_{l}$ being the Christoffel symbol. There are two things to note in this redefinition of the fields. First, $\{\mathcal{Q}, B_i\}=\{\mathcal{Q}, \{\mathcal{Q}, \bar\chi_i\}\}=0$ automatically because of the nilpotency of $\mathcal Q$. Second, the measure in the pathintegral in unchanged.

In terms of the new fields, the model is a W-TFT
\begin{subequations}
\label{WhiteNoisemodel}
\begin{eqnarray}
\langle\langle 1 \rangle\rangle = \iint_{[\Phi]}e^{i\{\mathcal{Q},\Theta\}},
\end{eqnarray}
with
\begin{eqnarray}
\Theta = \int_{R} \bar\chi_i\left(\partial_t\varphi^i + A^i + i g^{ij}(B_j + \bar\chi_k\Gamma^k_{jl}\chi^l) /2\right),\label{GaugeFermionReduced}\\
\{\mathcal Q,X\} = \int_{R}(\chi^i(R)\hat\delta_{\varphi^i(R)}
+B_i(R)\hat\delta_{\bar\chi_i(R)})X.\label{whiteQ}
\end{eqnarray}
\end{subequations}

We purposely used the transformed SDE method to quantize the model in order to make direct connection with the discussion in the previous section. In fact, model (\ref{WhiteNoisemodel}) can be obtained with less effort (with no redefinition of the fields) - out integration of the noise in the model constructed from the original SDE (\ref{reducedSDE}):
\begin{eqnarray}
\langle\langle 1 \rangle\rangle = \iint_{[\tilde\xi \Phi]} e^{i \{\mathcal Q, \int_R \bar\chi_i(\partial_t \varphi^i + A^i - e^i_a\tilde \xi^a)\} - (\tilde \xi^a)^2/2},
\end{eqnarray}
with $\mathcal Q$ from Eq.(\ref{whiteQ}), leads directly to Eqs.(\ref{WhiteNoisemodel}).

For $d=0$, model (\ref{WhiteNoisemodel}) is topological quantum mechanics, \cite{Labastida} in which case $H=\Sigma$. In many cases, below we consider this situation. Higher dimensional theories with $d>0$ can be viewed as infinite-dimensional generalizations of topological quantum mechanics.

\subsection{Schr\"odinger picture}
In passing from the path-integral representation of the theory to the operator algebra representation, the partition function is rewritten as:
\begin{eqnarray}
\langle\langle 1 \rangle\rangle = \iint_{[\Phi]} e^{\int_{t}(\int_{x}(iB_i\partial_t\varphi^i - i\bar\chi_i\partial_t\chi^i) - H(\Phi))}.\label{PathIntergalToFokkerPlank}
\end{eqnarray}
From here, the composition laws for the operators follow in the standard manner
\begin{eqnarray}
[\hat \varphi^i(x), \hat B_j(x') ]_- = - [\hat \chi^i(x), \hat {\bar\chi}_j(x')]_+ = i \delta^i_j \delta(x-x'). \label{Commutator}
\end{eqnarray}
The subscripts denote commutation for bosonic fields and anticommutation for the ghosts. Other (anti-)commutators are zero.

We chose to work in the representation where $\varphi$'s and $\chi$'s are diagonal. There are two reasons for this choice. First, these fields are superparthners. Second, in this basis the N\"oether charge associated with the topological symmetry is the exterior derivative that has no explicit dependence on the metric, which emphasizes the topological nature of the model.

In this basis, Eq. (\ref{Commutator}) suggests
\begin{eqnarray}
\hat B_i(x) = -i \hat \delta_{\varphi^i(x)}, \hat {\bar\chi}_i(x) = - i \hat \delta_{\chi^i(x)}.\label{operators}
\end{eqnarray}
Wavefunctions, $|\Psi\rangle$, are functionals of $\varphi$'s and $\chi$'s only. We can formally Taylor expand in $\chi$'s:
\begin{eqnarray}
|\Psi\rangle &=& \sum\nolimits_{n=0}^{\infty} |\Psi\rangle^{(n)},\nonumber\\
|\Psi \rangle ^{(n)} &=&\Psi^{(n)}_{(x_1k_1)...(x_nk_n)}\ast(\chi^{k_1}(x_1)...\chi^{k_n}(x_n)).\label{Ket}
\end{eqnarray}
Here, the integration over $x$'s and summation over $k$'s is assumed, and  $\Psi^{(n)}_{(x_1k_1)...(x_nk_n)}$ are antisymmetric in the pairs of subscripts due to the anticommutation composition law for the ghosts. For this reason, $|\Psi\rangle^{(n)}$'s can be interpreted \cite{WittenForms} as forms from the exterior algebra of $H$ which thus is the Hilbert space, $\mathcal H$, of the model.

In Eq.(\ref{Ket}), $\ast$ denotes the Hedge star, which is introduced for convenience. Within this definition, the probability density, which is a form of maximal degree, corresponds to $|\Psi\rangle^{(0)}$ as in Eq.(\ref{ProbabilityDemsity}) below.

The dynamical equation governing the time-evolution follows directly from the path-integral formulation in Eq.(\ref{PathIntergalToFokkerPlank}) and is the generalized Fokker-Plank equation:
\begin{subequations}
\label{FokkerPlankEq}
\begin{eqnarray}
\partial_t |\Psi \rangle= - \hat H |\Psi\rangle.
\end{eqnarray}
Its time reversed version for bra's, $\langle\Psi |\equiv(\ast |\Psi\rangle  )^*$, that are also forms on $H$, is
\begin{eqnarray}
\partial_t \langle\Psi| = \langle\Psi| \hat H^\dagger.\label{FPForBra}
\end{eqnarray}
\end{subequations}
The Hamitlonian in Eqs.(\ref{FokkerPlankEq}) is the generalized Fokker-Plank Hamiltonian. It's explicit form can be derived straightforwardly by the bi-graded symmetrization of its path-integral expression and with the use of Eq.(\ref{operators}). The result is well-documented in the Literature \cite{Frenkel} and is know to have a form of a N=2 (pseudo-)supersymmetric (pseudo-)Hermitian model: \cite{Mostafa}
\begin{subequations}
\begin{eqnarray}
\hat H = [\hat Q, \hat {\bar Q} ]_+/2 = - \nabla^2 /2 - \mathcal{L}_{A}. \label{FPHamiltonian}
\end{eqnarray}
Here
\begin{eqnarray}
\hat Q = \int_x \chi^i(x) \hat \delta_{\varphi^i(x)},\label{DeRahm}
\end{eqnarray}
is the conserved N\"oether charge associated with $\mathcal Q$-symmetry. Operator $\hat Q$ is nothing else but the exterior derivative on $\mathcal{H}$. Besides $\hat Q$, the Hamiltonian also conserves the number of ghosts given by the operator:
\begin{eqnarray}
\hat F=\int_x\chi^i(x)\hat\delta_{\chi^i(x)}.
\end{eqnarray}
In Eq.(\ref{FPHamiltonian}),
\begin{eqnarray}
\hat {\bar Q}= \hat Q^\dagger - 2 \int_x A^i(x)\hat\delta_{\chi^i (x)},\label{PseudoAdjoint}
\end{eqnarray}
is what could be called the (pseudo-)conjugate supercharge with
\begin{eqnarray}
\hat Q^\dagger = - \int_x \hat \delta_{\chi^i(x)}g^{ij}(\hat \delta_{\varphi^j(x)}-\chi^k\Gamma^l_{jk}\hat\delta_{\chi^l(x)}),
\end{eqnarray}
\end{subequations}
being the adjoint of the exterior derivative.

In Eq.(\ref{FPHamiltonian}), $-\nabla^2 = [\hat Q, \hat Q^\dagger]_+$ is the Laplace-Beltrami operator, explicit form of which is provided by the Weitzenb\"ock formula, and $\mathcal{L}_A = [\hat Q, \int_x A^i(x)\hat\delta_{\chi^i (x)}]_+$ is the Lie derivative along $A$.

We would also like to note, that formally the model is not a N=2 pseudo-supersymmetric in general for that reason that $\hat{\bar Q}$ is not nilpotent, $\hat{\bar Q}^2\ne0$, and it is not commutative with the Hamiltonian, $[\hat H, \hat{\bar Q}]_-\neq 0$.

\subsection{Connection to conventional Fokker-Plank equation}
\label{ConvFokkerPlank}
Because number of ghosts is the integral of motion, the time evolution does not mix wavefunctions of different ghost number. This, in particular, suggests that the time evolution of the wavefunction of the trivial (or rather maximal) ghost content:
\begin{eqnarray}
|\Psi\rangle^{(0)}(\varphi,t) \equiv \ast ( P(\varphi,t) ), \label{ProbabilityDemsity}
\end{eqnarray}
depends only on itself:
\begin{eqnarray}
\partial_t P = -\hat H^\text{cnv} P,\label{FP00}
\end{eqnarray}
where ($g^{1/2}$'s come from the Hedge star in Eq.(\ref{ProbabilityDemsity})):
\begin{eqnarray}
\hat H^\text{cnv} = \int_x g^{-1/2}\hat \delta_{\varphi^i(x)}g^{1/2}\left( - g^{ij} \hat \delta_{\varphi^j(x)}/2 - A^i(x)\right),\label{CnvFPHam}
\end{eqnarray}
is the conventional Fokker-Plank operator acting on the probability density, $P$.

If from a physical point of view, the probability density is a function(al), $P$, from the topological point of view, it is a form of the maximal degree as it is in Eq.(\ref{ProbabilityDemsity}). Notably, in stochastic quantization the probability density is not a "square" of a wave-function (of trivial ghost content) but rather the wave-function itself. The corresponding bra is actually the cycle over the entire $H$. This unconventional meaning of wave-functions is the direct consequence of the non-Hermitianity of the model that we discuss below

\subsection{Meaning of wavefunctions and $\mathcal Q$-symmetry}

In relation to the physical meaning of P, it is worth to briefly discuss the physical meaning of wavefunctions with non-trivial ghost content. In literature, $|\Psi\rangle^{(1)}$ is interpreted as current.\cite{currents} \footnote{Yet another possibility is to think of wave-functions of non-trivial ghost content as of conditional probabilities. We are not going to pursue this alternative interpretation in this paper, however.} It represents the flow of probabilities. This interpretation can be supported within the TFT formalism in the following manner. Consider topological quantum mechanics with $d$-dimensional phase space. Consider also Stock's theorem:
  \begin{eqnarray*}
  \int_{\partial\Omega}|\Psi\rangle^{(1)} = \int_{\Omega} (\hat Q|\Psi\rangle^{(1)}),
  \end{eqnarray*}
where $\Omega$ is a (fixed, time-independent) $d$-dimensional part of the phase space, $\partial\Omega$ is its ($d-1$)-dimensional boundary, and $|\Psi\rangle^{(1)}$ is any wavefunction with $d-1$ ghosts. We know that $(\hat Q|\Psi\rangle^{(1)})$ has the meaning of probability density so that $|\Psi\rangle^{(1)}$ must have the meaning of the  probability flow though $\partial\Omega$. Inductively, wavefunctions  of even more non-trivial ghost content must have the meaning of currents of currents of probability etc. Within this interpretation of wavefunctions, the model complies with the continuous higher-dimensional version of Kirchhoff's law. \cite{Kirchhoff} Then, operator from Eq.(26d) must be identified as conductance (or rather as conductance over capacitance).

The previous Stock's equality must hold at any moment of time and thus  it shows that $|\Psi\rangle^{(1)}$ and $\hat Q|\Psi\rangle^{(1)}$ evolve in time equivalently. Time evolution for both is given by Fokker-Plank equation so that the Fokker-Plank Hamiltonian must be commutative with $\hat Q$, which is the N\"other charge of $\mathcal Q$-symmetry. In other words, if a theory describes a dynamical model in terms of probability, currents of probabilities etc. it must possess topological supersymmetry in order to be consistent with Stock's theorem.

This meaning of wavefunctions also suggests that statistical description (in terms of probability density) of stochastic models is not applicable when $\mathcal Q$-symmetry is spontaneously broken. Indeed, $\hat Q|\text{ground}\rangle\ne0$ automatically suggests that the ground state is not a (pure) probability density since $\hat Q|\Psi\rangle^{(0)}=0$ for any $|\Psi\rangle^{(0)}$.

Wavefunctions with non-trivial ghost content may also be given the alternative interpretation of conditional probability densities. Even though there are a few physical reasonings that support this interpretation, we will not pursue this line of thinking in this paper.

\subsection{Non-Hermitianity}
\label{NonHermitianity}
There is a natural composition for bra's and ket's:
\begin{eqnarray}
\langle\Psi_1 |\Psi_2 \rangle = \iint_{H} (\ast \Psi_1)^* \wedge \Psi_2. \label{BraKetComposition}
\end{eqnarray}
In Hermitian/unitary models, Eq.(\ref{BraKetComposition}) is the metric in the Hilbert space, $\mathcal H$. In models under consideration, however, Hamiltonians are not Hermitian and the metric is not the one provided by Eq.(\ref{BraKetComposition}). To see how non-trivial metric shows up in non-Hermitian models, we address here the spectrum and eigenstates of the model.

It is convenient to think of $\hat H$ as of an infinite dimensional matrix with real entries. Its spectrum consists of either real eigenvalues or pairs of complex conjugate eigenvalues. If complex conjugate pairs exist, the model is said to be pseudo-Hermitian, while if the spectrum is purely real the model is said to be quasi-Hermitian. \cite{Mostafa}

For pseudo-Hermitian Hamitlonians, there exist a bi-orthogonal basis in the Hilbert space:
\begin{eqnarray}
\hat H |n\rangle\rangle = \mathcal{E}_n |n\rangle\rangle, \langle\langle n| \hat H = \langle\langle n| \mathcal{E}_n.\label{BiOrthBasis}
\end{eqnarray}
Here, $|n \rangle\rangle \equiv | n \rangle$ are merely the eigenstates of $H$, while $\langle\langle n | = \langle m |\hat \eta_{mn}$ are related to $\langle n|$'s through some non-trivial metric of the Hilbert space such that,
\begin{eqnarray}
\hat\eta\hat H\hat\eta^{-1}=\hat H^\dagger.
\end{eqnarray}
The bi-orthogonal basis is complete:
\begin{eqnarray}
\langle\langle n|m\rangle\rangle = \delta_{nm},
\hat 1_\mathcal {H} = \sum\nolimits_{n}|n\rangle\rangle\langle\langle n|,\label{Unity}
\end{eqnarray}
where $\hat 1_\mathcal{H}$ is the unity operator on $\mathcal H$. The metric mixes the "original" eigenstates with complex conjugate eigenvalues because:
\begin{eqnarray}
\langle n|\hat \eta \hat H = \langle n|\hat \eta \mathcal{E}_n^*.\label{timereversal}
\end{eqnarray}
The pairs of states with complex conjugate eigenvalues are time-reversal companions. This means that time-reversal operation acts non-trivially on such states.

The dynamical equation for a generic $\langle\langle \Psi |\equiv\langle \Psi|\hat \eta$ is
\begin{eqnarray}
\partial_t \langle\langle \Psi| = \langle\langle \Psi| \hat H,
\end{eqnarray}
instead of Eq.(\ref{FPForBra}).

In the literature on W-TFTs, pathintegrals as the one in Eq.(\ref{PathIntergalToFokkerPlank}), are often called Euclidian. This identification has a hidden danger in the context of dynamical systems. It may sound like the "actual" (or physical) temporal evolution is governed not by Eq.(\ref{FokkerPlankEq}) but by the Schr\"odinger equation related to Eq.(\ref{FokkerPlankEq}) through Wick rotation of time. The point here is that the time in Eq.(\ref{PathIntergalToFokkerPlank}) is the actual time of the SDE and the fundamental dynamical equation is the Fokker-Plank equation as it should. Therefore, apart from the subtle question of the validity of Wick rotation in pseudo-Hermitian models, this line of thinking may lead to an accidental "switching" of the physical meanings of the real and imaginary parts of the eigenvalues of the Hamiltonian.

To separate the real and imaginary parts of the eigenvalues we introduce:
\begin{eqnarray}
\mathcal{E}_n = \Gamma_n + iE_n.
\end{eqnarray}
In stochastic quantization, $\Gamma$'s are the attenuation rates (inverse lifetimes of eigestates), while $E$'s are quantum mechanical energies (inverse periods of closed classical trajectories associated with the eigenstates). This in particular suggests that stochastic quantization has a subtle difference with quantum dynamics. The role of the conventional kinetic term in quantum systems ($\sim - \Delta/2$) is to set the lower limit on the energies thus ensuring the existence of a ground state. In stochastic quantization, the kinetic term sets the lower limit on the attenuation rates thus caring more about the "stability" of the dynamics.

From physical reasonings it follows that the lower limit on $\Gamma$'s must be zero. Indeed, the states with $\Gamma<0$ would grow infinitely as time flows, thus indicating some sort of instability in the pathintegral representation of an SDE. On the other hand, the pathintegral represents actually the stochastic noise, which does not have any instabilities. Thus, states with $\Gamma_n<0$ must not exist. On the other hand, the partition function (which is again that of the noise) can not vanish in the $T\to\infty$ limit and thus must possess at least one "physical" state (see below) with $\Gamma$ being exactly zero.

The above physical reasoning about the possible generic form of the spectrum has its counterpart in the dynamical systems theory (see, \emph{e.g.}, Sec. III of Ref.[\onlinecite{Spectrum}]). There, the trace of the evolution operator plays the similar role as the partition function in the W-TFT approach. Perron-Frobenius operator is the analogue of the Fokker-Plank operator of the W-TFT. The zero-eigenvalue state, which must always exists, is called "ergodic" zero. All the other eigenstates must obey $\Gamma\ge0$. Complex eigenvalues also come in conjugate pairs and are called Ruelle-Pollicott resonances.

\subsection{Physical and ground states}
\label{PathIntegralSection}
Let us now turn back to the path-integral representation of the theory. Consider the expectation value of some observable, $\mathcal O$:
\begin{eqnarray}
\langle\langle\mathcal{O}\rangle\rangle = \iint_{[\Phi]} \mathcal{O}e^{i\{\mathcal{Q},\Theta\}}.\label{Expectation}
\end{eqnarray}
We are considering the equilibrium case, so that all the fields have periodic boundary conditions on a large temporal circle $t\in[T/2,-T/2]$. By integrating out the fields $B_i$'s and $\bar\chi$'s one obtains Eq.(\ref{Expectation}) in the operator algebra representation:
\begin{eqnarray}\nonumber
\langle\langle\mathcal{O}\rangle\rangle =\text{Tr}_\mathcal{H} (-1)^{\hat F}\mathcal{T}[\hat{\mathcal{O}_S}\hat U],
\end{eqnarray}
where $\hat{\mathcal{O}}_S$ is the observable in Schr\"odinger picture, $\hat U=e^{-\int_{t=-T/2}^{T/2}\hat H(t)}$ is the time evolution operator, and $\mathcal{T}$ denotes chronological ordering. In the previous formula, $(-1)^{\hat F}$ shows up because of the periodic boundary conditions for ghosts (see, \emph{e.g.}, Ref. [\onlinecite{book}]). In "normal" case of anti-periodic boundary conditions for fermionic fields, this factor does not exist.

One can now plug the unity from Eq.(\ref{Unity}) at temporal infinity, $\pm T/2$, and rewrite Eq.(\ref{Expectation}) as:
\begin{eqnarray}\nonumber
\langle\langle\mathcal{O}\rangle\rangle =\sum_n (-1)^{F_n}\langle\langle n|\mathcal{T}[\hat{\mathcal{O}}_S\hat U]|n\rangle\rangle,
\end{eqnarray}
The operator of temporal evolution acts on bra's and ket's in a simple manner: $\hat U|n\rangle\rangle = e^{-\mathcal{E}_nT}|n\rangle\rangle, \langle\langle n|\hat U = \langle\langle n| e^{-\mathcal{E}_n T}$. Using this, we further rewrite Eq.(\ref{Expectation}) as:
\begin{eqnarray}
\langle\langle\mathcal{O}\rangle\rangle =\sum_n (-1)^{F_n}e^{-\mathcal{E}_n T}\langle\langle n|\mathcal{T}[\hat{\mathcal{O}}_H]
|n\rangle\rangle.\label{IntermediateAverage1}
\end{eqnarray}
Here, $\mathcal{T}[\hat{\mathcal{O}}_H] \equiv \hat U^{-1}\mathcal{T}[\hat{\mathcal{O}}_S \hat U]$ is the chronologically ordered $\hat{\mathcal O}$ in the Heisenberg representation. Diagonal matrix elements of $\mathcal{T}[\hat{\mathcal{O}}_H]$ do not depend on $T$. For example, if the operator is a two-time correlator, $\hat{\mathcal{O}} = \hat X(t)  \hat Y(t'), t>t'$, its diagonal matrix element is
\begin{eqnarray}
\langle\langle n|\mathcal{T}[\hat{\mathcal{O}}_H]|n\rangle\rangle = \sum_{m} X_{nm} Y_{mn} e^{-(\mathcal{E}_{m} - \mathcal{E}_{n})(t-t')}.
\end{eqnarray}
Now it follows that in the $T\to 0$ limit, only states with $\Gamma_n=0$ contribute to Eq.(\ref{Expectation}). Only these states can be considered physical ones as only they survive the infinitely long Fokker-Plank evolution and thus can appear as the out-states in the scattering matrix.
\begin{figure}
\includegraphics[width=8.0cm,height=4.2cm]{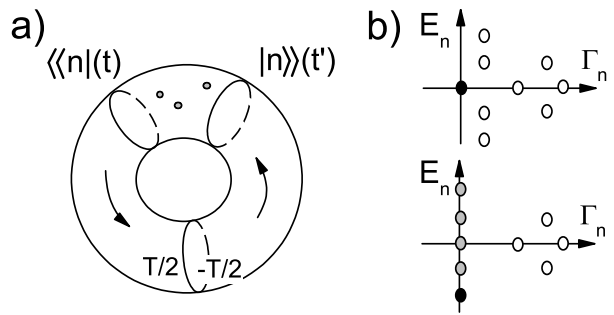}
\caption{ \label{Figure2}{\bf (a)} Graphical representation of the manipulations with the path-integral in Sec.(\ref{PathIntegralSection}). Curly arrows represent the direction of time. The infinitely far future and past are integrated out around some finite time domain $t>t'$, of some observable denoted as a collection of dots. This evolves the eigenstates according to $\langle\langle n|(t)=\langle\langle n|e^{-\mathcal{E}_n(T/2-t)}$ and $|n\rangle\rangle(t')= e^{-\mathcal{E}_n(t'-(-T/2))}|n\rangle\rangle$. In the physical limit, $T\to\infty$, only the physical states with zero real parts of their eigenvalues, $\mathcal{E}_n$, survive. {\bf (b)} Qualitative comparison of the generalized Fokker-Plank Hamiltonian spectra in models with unbroken (top) and spontaneously broken (bottom) $\mathcal Q$-symmetry. $E_n=\text{Im}\mathcal{E}_n$ and $\Gamma_n=\text{Re}\mathcal{E}_n$ are the imaginary and the real parts of the eigenvalues of the Fokker-Plank Hamiltonian, $\mathcal{E}_n$. Unphysical states with $\Gamma_n>0$, physical states with $\Gamma_n=0$, and physical ground states with minimal $E_n$ are represented respectively as hollow circles, circles with grey filling, and black circles. $\mathcal Q$-symmetry is broken on states with $\mathcal{E}_n\ne0$ and time-reversal symmetry must be broken on states with $E_n\ne0$ (Ruelle-Pollicott resonances). In case of unbroken $\mathcal Q$-symmetry, the only physical states are the ground states. }
\end{figure}

In fact, $\Gamma_n=0$ should be looked upon only as a necessary condition for a state to be a physical. Other conditions may apply. One of the examples of such conditions exists in quantum field theories where states that have negative norm are viewed as non-physical. This condition does not apply, however, to W-TFTs (see, \emph{e.g.}, Sec. 3.6.3. of Ref.[\onlinecite{TFT}]). Our temporal understanding of the physical meaning of the negative-norm states (those with odd number of ghosts) is through the concept of negative probabilities discussed in Sec. \ref{NegativeProb}.

We can adopt yet another condition for physicality of a state by requiring that the norm of a state is non-zero. We will encounter such possibility in the next subsection. At this moment, we are not aware of any other conditions for states to be considered physical. For this reason, allow us to call all states with $\Gamma_n=0$ physical ones. The Hilbert space $\mathcal H$ is split now as:
\begin{eqnarray}
\mathcal{H} = \mathcal{H}^p \oplus \mathcal{H}^u,
\end{eqnarray}
where $\mathcal{H}^p$ is spanned by all the physical states, and $\mathcal{H}^u$ is spanned by the non-physical states with $\Gamma_n>0$.

In the $T\to\infty$ limit, Eq.(\ref{IntermediateAverage1}) takes the form
\begin{eqnarray}
\langle\langle\mathcal{O}\rangle\rangle =\sum_{|n\rangle\rangle\in\mathcal{H}^p} (-1)^{F_n}e^{-i E_{n}T}\langle\langle n|\mathcal{T}[\hat{\mathcal{O}}_H ]|n\rangle\rangle.\label{IntermediateAverage}
\end{eqnarray}
One notes that there is an oscillating factor, $e^{-iE_nT}$, that suppresses the summation (or rather the integration in the higher dimensional theories) over the physical states. In the spirit of quantum field theories, one can formally Wick rotate "a little" the parameter $T\to T(1-i 0^+)$. Again, in the $T\to\infty$ limit this will eliminate all the physical states from Eq.(\ref{IntermediateAverage}) except those with the lowest $E_g = \min_{n\in\mathcal{H}^p} E_n$. These physical states could be called ground states. In result we get:
\begin{eqnarray}
\langle\langle\mathcal{O}\rangle\rangle \sim \sum_{|n\rangle\rangle\in\mathcal{H}^g} (-1)^{F_n}\langle\langle n|\mathcal{T}[\hat{\mathcal{O}}_H ]|n\rangle\rangle.\label{IntermediateAverage2}
\end{eqnarray}
where $\mathcal{H}^g$ is the part of the Hilbert space spanned by the ground states.

The meaning of the similarity sign in the last expression is twofold. First, we dropped the factor $e^{-iE_gT}$. The second meaning is more fundamental. The point is that for $\mathcal Q$-symmetry broken cases, Eq.(\ref{IntermediateAverage2}) may not be a good approximation for some observables. An example of such an observable is the unity operator. For $\mathcal{O} = \hat 1_{\mathcal{H}}$, the average is the Witten index
\begin{eqnarray}
\langle\langle 1 \rangle\rangle = \sum_n (-1)^{F_n} e^{-T\mathcal{E}_n},\label{WittenIndex}
\end{eqnarray}
which is a well known fact that the partition function of a W-TFT is the Witten index. The only difference in the stochastic quantization case is that $\mathcal{E}_n$'s can be complex.

As it will be discussed in the next subsection, all the eigenstates with non-zero eigenvalues break $\mathcal Q$-symmetry and thus must come in bosonic-fermionic pairs, who's contribution must cancel each other out of the partition function. Therefore, only the zero-eigenvalue states must contribute. In result, in topological quantum mechanics Witten index equals Euler characteristic of the target manifold, which in general is not zero. This must be true even for cases with broken $\mathcal Q$-symmetry - after all, the partition function is that of the noise and thus it must not depend on the flow. On the other hand, if we used Eq.(\ref{IntermediateAverage2}) for the calculation of the partition function in this cases, we would get zero. Consequently, it is Eq.(\ref{IntermediateAverage}) that must be used for a general observable.

Turning back to the path-integral representation brings Eq.(\ref{IntermediateAverage}) to a "more" field-theoretic form:
\begin{eqnarray}
\langle\langle\mathcal{O}\rangle\rangle = \sum_{|n\rangle\rangle\in\mathcal{H}^p} (-1)^{F_n}\iint_{[\Phi]}\langle\langle n| \mathcal{O} e^{i\{\mathcal{Q},\Theta\}}|n\rangle\rangle,\label{ObservableFinal}
\end{eqnarray}
that will be helpful in the next subsection.

The purpose of the textbook-level exercise in this subsection was twofold. First, it showed that the non-trivial metric of the Hilbert space of a pseudo-Hermitian model is automatically incorporated into the pathintegral representation. \cite{PathIntegralMetric} Second, it revealed the essence of the unphysical, physical, and ground states of the model.

\subsection{$\mathcal{Q}$- and time reversal symmetries of eigenstates}
\label{QsymmetryStates}

$\mathcal Q$-symmetry is pertinent to path-integral representations of all dynamical systems. A generic eigenstate, however, does not possess the symmetry of the Hamiltonian. If this is true for a ground state, it is said that the symmetry is broken spontaneously. Therefore, in order to judge if $\mathcal Q$-symmetry is broken or not, it is important to understand which eigenstates break it. This is the primary goal of this subsection.

The most general criterion of that an eigenstate does not break some continuous symmetry is that the expectation value of any observable, which is a result of acting by an infinitesimal operator of this symmetry on something, is zero. For $\mathcal{Q}$-symmetry, this criterion takes the following form (c.f., Eq.(\ref{ObservableFinal})):
\begin{eqnarray}
\iint_{\Phi}\langle\langle n| \{\mathcal {Q}, X\} e^{i\{\mathcal{Q},\Theta\}} |n\rangle\rangle = 0,\label{UnbrokenCondition1}
\end{eqnarray}
for any $X$. Using integration by parts with respect to $\mathcal Q$ and $\{\mathcal{Q}, e^{i\{\mathcal{Q},\Theta\}}\}=0$, Eq.(\ref{UnbrokenCondition1}) reduces to:
\begin{subequations}
\label{UnbrokenCondition}
\begin{eqnarray}
\{\mathcal {Q}, |n\rangle\rangle\}  =  \hat Q |n\rangle\rangle =0, \\
\pm\{\mathcal {Q}, \langle\langle n|\}  = \langle\langle n|\hat Q = 0.
\end{eqnarray}
\end{subequations}
Here we used the fact that bra's and ket's depend only on $\varphi, \chi$ so that $\mathcal Q$ operator from Eq.(\ref{Qsymmetry}) becomes $\hat Q$ operator from Eq.(\ref{DeRahm}). The choice of sign in the second line depends on the ghost degree of $\langle\langle n|$.

In combination with Eqs.(\ref{BiOrthBasis}) and (\ref{FPHamiltonian}), Eqs.(\ref{UnbrokenCondition}) also suggests that:
\begin{subequations}
\label{ScndUnbrokenCond}
\begin{eqnarray}
\hat Q\hat{\bar Q} |n\rangle\rangle = 2 \mathcal{E}_n |n\rangle\rangle,\\
\langle\langle n| \hat{\bar Q} \hat Q  = 2 \mathcal{E}_n \langle\langle n|,
\end{eqnarray}
\end{subequations}
with $\hat{\bar Q}$ from Eq.(\ref{PseudoAdjoint}). Accordingly, bra's and ket's of eigenstates that do not break $\mathcal Q$-symmetry and at the same time have $\mathcal{E}_n\ne0$ are $\mathcal Q$-exact: $|n\rangle\rangle = \hat Q |x_n\rangle\rangle, \langle\langle n| = \langle\langle x_n|\hat Q$, where $|x_n\rangle\rangle = \hat{\bar Q} |n\rangle\rangle/(2\mathcal{E}_n), \langle\langle x_n| = \langle\langle n|\hat{\bar Q}/(2\mathcal{E}_n)$. This, in turn, suggests that $\langle\langle n|n\rangle\rangle=\langle\langle x_n|\hat Q^2|x_n\rangle\rangle=0$. In other words, if $\mathcal Q$-symmetry is not broken by an eigenstate with non-zero eigenvalue, then it has zero norm. This situation is exotic. If such situation occurs, some measures must be undertaken to get rid of this state.

The temporary conclusion is that all the eigenstates with non-zero eigenvalue break $\mathcal Q$-symmetry. From this it follows that the $\mathcal Q$-symmetry is unbroken spontaneously if the only physical states (those with $\Gamma_n=0$) are those with $E_n=0$. In other words, the only physical states are the ground states that satisfy ($n \in \mathcal{H}^g = \mathcal{H}^p$)
\begin{subequations}
\label{RealConditions}
\begin{eqnarray}
\hat Q |n\rangle\rangle = 0, \langle\langle n|\hat Q= 0, \\
\hat Q\hat{\bar Q} |n\rangle\rangle = 0, \langle\langle n| \hat{\bar Q} \hat Q =  0.
\end{eqnarray}
\end{subequations}
These conditions are very similar to those of the unbroken N=2 pseudo-supersymmetry \cite{Mostafa} with the only difference that the second line is weaker than its N=2 pseudo-supersymmetric version: $\hat{\bar Q} |n\rangle\rangle = 0, \langle\langle n| \hat{\bar Q} =  0$.

The comparison of spectra of models with broken and unbroken $\mathcal Q$-symmetry is given in Fig.\ref{Figure2}b. In essence, the spontaneous $\mathcal Q$-symmetry breaking can be interpreted as the condensation of (a branch of) Ruelle-Pollicot resonances (\emph{i.e.}, eigenstates with imaginary eigenvalues) into the physical part of the Hilbert space.

Speaking of time-reversal symmetry, as it follows from Eq.(\ref{timereversal}), each Ruelle-Pollicott resonance, either condensed or not, is not the time-reversal companion of itself. This suggests that time-reversal symmetry must be broken on these states. It then follows (see Fig.\ref{Figure2}b) that the spontaneous breakdown of $\mathcal Q$-symmetry must always be accompanied by the spontaneous breakdown of time-reversal symmetry. This line of thinking is very similar to that about the spontaneous breakdown of N=2 pseudo-supersymmetry in Ref.[\onlinecite{SinhaRoy}], where it was shown that it must always come together with the spontaneous breaking of time-reversal symmetry.

\section{Deterministic limit: ground states}
\label{VacuaDetermin}
In this section, we would like to discuss the form of the ground states of topological quantum mechanics ((0+1) theory) in the deterministic limit for cases of unbroken $\mathcal Q$-symmetry. The discussion will provide a route to understanding the origin of the spontaneous $\mathcal Q$-symmetry breaking on the mean-field level, which happens in chaotic models (see Sec.\ref{Chaos}).

In the deterministic limit, when the Fokker-Plank Hamiltonian reduces to the Lie derivative, the ground states, $|0\rangle\rangle$, are closed states ($\hat Q|0\rangle\rangle=0$) that satisfy
\begin{eqnarray}
\mathcal{L}_A |0\rangle\rangle = 0.\label{LieDerivative}
\end{eqnarray}
Lie derivative, however, is not a bounded operator. Therefore, a more appropriate route is to perform a one-loop analysis of a model with the metric:
\begin{eqnarray}
g^{ij}=\epsilon g_0^{ij}, \label{deterministicnoise}
\end{eqnarray}
and then send the "intensity" of the noise to its deterministic limit, $\epsilon\to0$. The exact form of $g_0^{ij}$ must not matter as long as it is decent, \emph{e.g.}, positive definite.

Let us first consider an isolated critical point, $\varphi_0,A^i(\varphi_0)=0$. The one-loop partition function of the tower of the one-loop eigenstates around $\varphi_0$ is:
\begin{eqnarray}
\langle\langle |1|\rangle\rangle^\text{1-loop}_{\varphi_0} &=& \iint e^{i\{\mathcal{Q},\Theta(\varphi_0)\}}\nonumber\\
&=&\sum\nolimits_n(-1)^{F_n}e^{-T\mathcal{E}_n} = (-1)^{\Delta_{\varphi_0}}.\label{OneLoopContr}
\end{eqnarray}
Here the gauge fermion generating the corresponding gaussian action is $\Theta(\varphi_0) = \int_{t=-T/2}^{T/2} \bar\chi_i(\partial_t \delta\varphi^i+A^i_j\delta\varphi^j+i\epsilon g^{ij}_0B_j)$, $\delta\varphi^i=\varphi^i-\varphi^i_0$, $n$ enumerates the one-loop eigenstates, $A^i_j = \partial_{\varphi^j} A^i(\varphi_0)$ so that $A^i(\varphi) \approx A^i_j\delta\varphi^j$.

The last equality sign in Eq.(\ref{OneLoopContr}) follows from the famous cancelation of bosonic and fermionic determinants in W-TFT's (see \emph{e.g.}, Ref.[\onlinecite{Labastida}]), with $\Delta_{\varphi_0}$ being the number of negative real eigenvalues of $A^i_j$.

Eq.(\ref{OneLoopContr}) says that there exist (at least one) one-loop state with zero eigenvalue of the Hamiltonian. The zero-energy state is either bosonic or fermionic depending on $\Delta_{\varphi_0}$. All the states with non-zero eigenvalues come in bosonic-fermionic pairs so that their contribution cancels out from the partition function.

In this subsection, we call zero-energy state a ground state. As long as we are talking about the unbroken $\mathcal Q$-symmetry case, this is reasonable.

We consider what could be called the minimal "non-potential" generalization of a potential flow near $\varphi_0$. We assume that there exist local orthogonal coordinates, $\tilde\varphi^i=O^{i}_{j}\delta\varphi^j$, in which $A^{i}_{j}$ is block-diagonal with diagonal elements being either $1\times 1$ or $2\times 2$ matrices. One-dimensional blocks corresponds to real eigenvalues of $A^{i}_{j}$ and to the directions, in which the flow is locally potential. The two-dimensional blocks correspond to pairs of complex conjugate eigenvalues that represent, \emph{e.g.}, sinks. All eigenvalues have non-zero real part. Otherwise, the critical point is not isolated and belongs to a higher-dimensional invariant manifold.

In these coordinates, the directions corresponding to different eigenvalues of $A^i_j$ become independent (we chose $g_0^{ij}$ that does not mix coordinates of different diagonal blocks of $A^i_j$). The ground state will be the wedge product of the ground states in each of the coordinates or the pairs of coordinates:
\begin{eqnarray}
|\varphi_0\rangle\rangle = \prod\nolimits_\alpha ^{(\wedge)}|\alpha\rangle\rangle \prod\nolimits_\beta^{(\wedge)} |\beta\rangle\rangle,\label{DecompositionKet}
\end{eqnarray}
where $\alpha$ and $\beta$ numerate respectively the coordinates of the single real eigenvalues of $A^i_j$ and the pairs of coordinates of the complex eigenvalues. Similar factorization holds for bra of the ground state.

We consider first a coordinate, $\tilde\varphi^\alpha\equiv\tilde\varphi$, corresponding to a real eigenvalue of $A^i_j$, $\lambda$. We chose metric $g_0^{\alpha\alpha}=1$. It is straightforward to verify that the ground state satisfying Eqs.(\ref{RealConditions}) with:
\begin{eqnarray}
\hat Q  = \chi \hat \partial_{\tilde\varphi}, \hat{\bar Q} = \hat \partial_{\chi}(-\epsilon \hat \partial_{\tilde\varphi} - 2 \lambda \tilde\varphi),
\end{eqnarray}
is
\begin{eqnarray}
|\alpha\rangle\rangle \sim
e^{-|\lambda|\tilde\varphi^2/\epsilon} \chi,
\langle\langle \alpha| \sim 1,\label{BraLangevin}
\end{eqnarray}
for a stable coordinate ($\lambda>0$) and $|\alpha\rangle\rangle \leftrightarrow \langle\langle \alpha|$ for an unstable coordinate ($\lambda<0$).

In the Hermitian basis of a Langevin model (see Eqs.(\ref{PsiTransfromBra}) and (\ref{PsiTransfromKet}) below) and with the corresponding superpotential, $W = \lambda\tilde\varphi^2/2\epsilon$, the wave-function localizes to the critical point
\begin{eqnarray}
|\alpha\rangle_\text{L}=\ast(\langle\alpha|_\text{L})^* \sim e^{-|\lambda|\tilde\varphi^2/(2\epsilon)} \left\{\begin{array}{lc}\chi,&\lambda>0,\\1, &\lambda<0,\end{array}\right.\label{ketWitten}
\end{eqnarray}
This is the well-known one-loop ground state in Witten models (see, \emph{e.g.}, Chap. 10.3.3 of Ref.[\onlinecite{book}]). The only difference here is that we are using the representation in which $\hat\chi$'s are diagonal instead of $\hat{\bar\chi}$'s, so that $\chi\leftrightarrow 1$ brings Eq.(\ref{ketWitten}) to its more conventional form.

Let us turn to two coordinates, $\tilde\varphi^\beta \equiv (\tilde\varphi^1,\tilde\varphi^2)$, with a complex conjugate pair of eigenvalues, $\lambda^\pm = \lambda'\pm\lambda''$. First, we note that the two coordinates contribute the factor $+1$ into Eq.(\ref{OneLoopContr}). This suggests that the ground state is bosonic, \emph{i.e.}, it has either two or zero ghosts. As in case of a single eigenvalue of $\hat A$, the first case corresponds to two stable coordinates ($\lambda'>0$) and the second case corresponds to two unstable coordinates ($\lambda'<0$). Let us consider the first case.

Without loss of generality:
\begin{eqnarray}
A^i_j =
\left(\begin{array}{cc}
\lambda' & \lambda'' e^{\kappa}\\
-\lambda''e^{-\kappa} & \lambda'
\end{array}\right),\label{AgeneralForm}
\end{eqnarray}
with $\kappa\geq 0$. We can now use the freedom in Eq.(\ref{deterministicnoise}) of choosing a metric and set
\begin{eqnarray}
\hat g_0 = \hat A^+ /\Lambda, \Lambda = (\tilde\lambda^-\tilde\lambda^+)^{1/2}.\label{Metricg0}
\end{eqnarray}
Here $\hat A^+ = (\hat A + \hat A^\text{T})/2$ and $\tilde \lambda^\pm = \lambda' \pm \lambda''\sinh\kappa$ are the eigenvalues of $\hat A^+$. At this, the condition shows up that $\hat A^+$ is positive definite. We assume it is satisfied.

It can be straightforwardly verified that:
\begin{eqnarray}
|\beta\rangle\rangle \sim e^{-\Lambda \tilde\varphi^2/\epsilon}\chi^1\chi^2, \langle\langle\beta| \sim 1, \label{2Ground}
\end{eqnarray}
with $\tilde\varphi^2 = (\tilde\varphi^1)^2+(\tilde\varphi^2)^2$, satisfies Eqs.(\ref{RealConditions}) with:
\begin{eqnarray}
\hat Q  = \chi^i \hat \partial_{\tilde\varphi^i}, \hat{\bar Q} = \hat \partial_{\chi^i}(-\epsilon g_0^{ij} \hat \partial_{\tilde\varphi^j} - 2 A^i_j \tilde\varphi^j),
\end{eqnarray}
where $i,j=1,2$, and $g_0$ and $A$ from Eqs.(\ref{Metricg0}) and (\ref{AgeneralForm}). In a similar way, for a pair of unstable coordinates one similarly gets the ground state with $|\beta\rangle\rangle \leftrightarrow \langle\langle\beta|$.

Eqs. (\ref{BraLangevin}) and (\ref{2Ground}) show that in the deterministic limit, $\epsilon\to0$, and under certain conditions, the one-loop bra and ket of the ground state of an isolated critical point localize and/or represent respectively stable and unstable manifolds, that intersect on this critical point. Bra's and ket's are forms in the transverse directions and have no coordinate dependence in the tangential directions. This picture must hold for more intricate flows of hyperbolic critical points. The coordinate dependence of the ket of the ground state must be related to Lyapunov function on the stable manifold. Such function must always exist on the stable manifold.

The closest concept in dynamical systems theory we managed to find in the Literature are discussed in Secs. 3.3-4. of Ref.[\onlinecite{GASPARD}]. There, the distributions that localize to unstable manifolds of critical points were identified as Gel'fand-Schwartz distributions.

The analysis can be generalized to cases when critical points are not isolated and form higher-dimensional critical manifolds, $M_c$ (Morse-Bott-type case). The analysis near each point of $M_c$ is split into the directions tangent and transverse to $M_c$. In tangent directions, the Hamiltonian is the Laplacian and ground states are harmonic forms, $|\omega_i\rangle\rangle$, $\triangle_{M_c}|\omega_i\rangle\rangle=0$. In transverse directions, if conditions are right, the ground state is the one discussed previously in the context of an isolated critical point, \emph{i.e.}, the ket, $|\mu\rangle\rangle$, is a volume form on stable manifold and is localized to the unstable manifold. Provided that $|\mu\rangle\rangle$'s at different points of $M_c$ can be glued together over the entire $M_c$, the ground states are $|i,M_c\rangle\rangle = |\omega_i\rangle\rangle\wedge|\mu\rangle\rangle$.

\begin{figure}
\includegraphics[width=8.0cm,height=3.5cm]{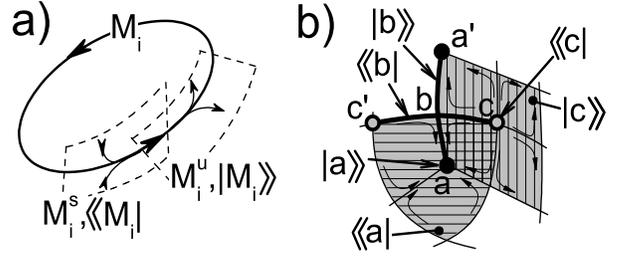}
\caption{\label{Figure3} {\bf (a)} In deterministic limit and under some general conditions on the flow, bra's, $\langle\langle M_i|$, and ket's, $|M_i\rangle\rangle$, of perturbative ground states around an invariant manifold, $M_i$, represent respectively local stable, $M^s_i$, and unstable, $M^u_i$, manifolds intersecting on $M_i$. {\bf (b)} An example of perturbative ground states for a potential-like flow. A stable critical point, $a$, a saddle, $b$, and an unstable critical point, $c$, are under consideration. $|a\rangle\rangle$ (black circle) is a maximal-degree form  that is localized to the critical point, while $\langle\langle a|$ (area with horizontal filling) is a constant form with no ghosts and represents the entire basin of attraction of this critical point. $|b\rangle\rangle$ (think curve $aba'$) and $\langle\langle b|$ (think curve $c'bc$) are forms representing respectively unstable and stable manifolds of $b$. They are forms in transverse directions. $|c\rangle\rangle$ (area with vertical filling) is a constant form with no ghosts representing the basin of repulsion of c, while $\langle\langle c|$ (hollow circle) is a maximal degree form localized at this critical point.}
\end{figure}

Similar situation occurs for perturbative ground states around invariant manifolds, $M_i$. Again, the analysis is split into tangent and transverse directions. In transverse directions, the ground state is the same as in Bott-Morse-type case, $|\mu\rangle\rangle$. In tangent directions, one can chose such metric $g_0$ from Eq.(\ref{deterministicnoise}) (if it exists) that $A$ is a Killing vector field on $M_i$. Then, the tangential ground states are the invariant harmonic forms also satisfying $\mathcal{L}_A|\omega_i\rangle\rangle=0$. The ground states again are as in the Bott-Morse case, $|i,M_i\rangle\rangle=|\omega_i\rangle\rangle \wedge |\mu\rangle\rangle$. A toy model of the simplest invariant manifold, a limit cycle, could be a topological quantum mechanics on a circle with constant $A$ considered in Sec.\ref{RefinedDefinitionofChaos} below. In this case, the ground states are from the cohomology of the circle, $|\omega_i\rangle\rangle=1,\chi$, \emph{i.e.}, $|\Psi\rangle\rangle^{(0)}_0$ and $|\Psi\rangle\rangle^{(1)}_0$ in notations of Sec.\ref{RefinedDefinitionofChaos}.

For the perturbative analysis, only the combinations/intersections of bra's and ket's are important. These combination are localized to invariant manifolds. However, bra's and ket's separately are not localized. In fact, perturbative analysis does not say how far bra and ket stretch away from the invariant manifold. This provides some sort of a freedom in the definition of perturbative ground states. One can use this freedom to incorporate the effect of instantons - classical solutions that connect different invariant manifolds. The point is that instantons provide natural boundaries to stable and unstable manifolds of a given invariant manifold. An example of such a definition of perturbative ground states is given in Fig.\ref{Figure3}b, where a potential-like flow in a two dimensional phase space is presented. There, the perturbative ground states are given for three critical points with indices zero (stable critical point, $a$), one (saddle, $b$), and two (unstable critical point, $c$). $|a\rangle\rangle$ is a maximal-degree form that is localized to the critical point, while $\langle\langle a|$ is a zero-ghost constant form that represents the entire basin of attraction of $a$. $|b\rangle\rangle$ and $\langle\langle b|$ are forms localized to respectively the unstable and stable manifolds of $b$ and they both have ghosts in the transverse directions. $|c\rangle\rangle$ is a constant form representing the "repulsion" basin of c, while $\langle\langle c|$ is a maximal degree form localized to this critical point.

As is seen, ket's of perturbative ground states represent the instantonic CW-complex of the phase space, while bra's represent the dual anti-instantonic CW-complex. It is also seen that $\hat Q$ operator acts on perturbative ground states just as a boundary operator would have acted on the corresponding CW-complex. For example:
\begin{eqnarray}
\hat Q |b\rangle\rangle = |a\rangle\rangle - |a'\rangle\rangle,\label{Morse-Witten}
\end{eqnarray}
or, similarly,
\begin{eqnarray}
\langle\langle b| \hat Q = \langle\langle c| - \langle\langle c'|.
\end{eqnarray}
In case of a potential flow with isolated critical points, perturbative ground states and equalities like the one in Eq.(\ref{Morse-Witten}) form Morse-Witten complex, cohomology of which must be equivalent to the homology of the instantonic CW-complex as follows from above discussion. Under certain conditions, methodology of the Morse-Bott complex can be generalized to non-potential flows. The proposed view on the perturbative ground states can be useful for such generalization.

From the global point of view, the perturbative ground states are not $\hat Q$-closed as is seen, \emph{e.g.}, from Eq.(\ref{Morse-Witten}). This, however, plays no role on the perturbative level since, \emph{e.g.},  $\langle\langle b|\hat Q |b\rangle\rangle =0$. In order to get the global ground states beyond the perturbative analysis, one must glue perturbative ground states to form globally defined stable and unstable manifolds, \emph{i.e.}, consider superposition of perturbative ground states that are globally $\mathcal Q$-closed.

In the dynamical systems theory, there are theorems that establish the conditions, under which the global topologically well-defined stable and unstable manifolds criss-crossing on invariant manifolds exist. It is natural to expect that these conditions are also necessary conditions for unbroken $\mathcal Q$-symmetry in the deterministic limit. Those may not be sufficient conditions, however. Establishing the conditions of unbroken $\mathcal Q$-symmetry in the deterministic limit in terms of the properties of the flow may turn out to be a non-trivial problem that we leave open. We would only like to mention here, that to our temporary and mostly intuitive understanding these conditions may have a lot to do with the integrability conditions. In terms of the spectrum of Fokker-Plank Hamiltonian, however, the necessary and sufficient conditions for unbroken $\mathcal Q$-symmetry look fairly clear (see Fig.\ref{Figure2}b).

As a closing part of this section, we would like to point out
the equivalence of Eq.(60) and the supersymmetry operator in
stochastic networks discussed, for example, in Ref.[\onlinecite{network}]. There,
the supersymmetry relates probability currents and probability
densities defined on a network. The direct analogy with our case
is as follows: the supersymmetry is the exterior derivative,
the network is the instantonic CW-complex, and, say,
$|b\rangle\rangle$ is the current between $|a\rangle\rangle$
and $|a'\rangle\rangle$.

\section{Unbroken $\mathcal Q$-symmetry}
\label{UnbrokenSymmetry}
\subsection{Langevin SDEs}
\label{LangevinSDEs}
The important class of models with unbroken $\mathcal Q$-symmetry is the Langevin SDEs. They represent purely dissipative dynamics. This is the most studied class of dynamical systems in the context of W-TFTs. The models are quasi-Hermitian and have a Hermitian representation known as Witten models. \cite{WittenForms} As we discuss in this subsection, $\mathcal Q$-symmetry is never broken for Langevin SDE's (for equilibrium dynamics) even in the presence of noise.

Langevin SDEs are those with a potential flow:
\begin{eqnarray}
A^i_\text{L}(x) = g^{ij}\hat\delta_{\varphi^i(x)}W,\label{LangevinDrift}
\end{eqnarray}
where $W$ is some functional called superpotential. All the eigevalues of the corresponding Fokker-Plank Hamiltonian are real. The Hamiltonian can be brought to a Hermitian form by the similarity transformation
\begin{subequations}
\label{Transformation}
\begin{eqnarray}
\hat H\to \hat H_\text{L} = \hat H_\text{L}^\dagger = \hat \eta^{1/2}_\text{L} \hat H \hat \eta^{-1/2}_\text{L},
\end{eqnarray}
where $\hat \eta_\text{L}=e^{2W}$ is the Hilbert space metric discussed in Sec.\ref{NonHermitianity}. The corresponding transformation of the Hilbert space and the supercharges is
\begin{eqnarray}
\langle\langle \Psi| &\to& \langle\Psi|_\text{L} = \langle\langle \Psi|\hat \eta_\text{L}^{-1/2},\label{PsiTransfromBra}\\
|\Psi\rangle\rangle &\to& |\Psi\rangle_\text{L} = \hat \eta^{1/2}_\text{L}|\Psi\rangle\rangle,\label{PsiTransfromKet}\\
\hat Q &\to& \hat Q_\text{L} = \hat Q + \int_x\chi^i(x)(\hat\delta_{\varphi^i(x)} W),\\
\hat{\bar Q} &\to& \hat {\bar Q}_\text{L} = \hat Q^\dagger - \int_x\hat\delta_{\chi^i(x)}g^{ij}(\hat\delta_{\varphi^j(x)} W).
\end{eqnarray}
\end{subequations}
It is easy to see in this hermitian form that the model has yet another nilpotent supersymmetry charge, $\hat{\bar Q}_\text{L}, \hat{\bar Q}_\text{L}^2=0, [\hat H, \hat{\bar Q}_\text{L}]_-=0$, which is the Hermitian conjugate and the time-reversal companion of $\hat Q_\text{L}=\hat {\bar Q}_\text{L}^\dagger$. The model became N=2 supersymmetric with the Hamiltonian
\begin{eqnarray}
2\hat H_\text{L} = [\hat Q_\text{L}, \hat Q^\dagger_\text{L}]_+ = \hat Q_1^2 = \hat Q_2^2,
\end{eqnarray}
where the two Hermitian supercharges are $\hat Q_1 = \hat Q_\text{L} + \hat {\bar Q}_\text{L}$ and $\hat Q_1 = i(\hat Q_\text{L} - \hat {\bar Q}_\text{L})$.

In this representation, all the eigenvalues are real and non-negative, $\mathcal{E}_n = \Gamma_n$:
\begin{eqnarray}
2\Gamma_n = |\hat Q_\text{L}|n\rangle_\text{L}|^2 + |\hat {\bar Q}_\text{L}|n\rangle_\text{L}|^2\geq0.
\end{eqnarray}
For ground states we have $\Gamma_n=0$ so that:
\begin{eqnarray}
\hat Q_\text{L}|n\rangle_\text{L}=0, \hat {\bar Q}_\text{L}|n\rangle_\text{L}=0.
\end{eqnarray}
Now it follows that $\mathcal Q$-symmetry, together with the N=2 supersymmetry, can not be broken if at least one physical ground state exist. Such ground state does always exist. It has the from of Eq.(\ref{ProbabilityDemsity}):
\begin{eqnarray}
|0\rangle\rangle=(\ast P_0), \langle\langle 0|= (\ast|0\rangle\rangle)^*\hat \eta_L = 1.\label{StationaryDistribution}
\end{eqnarray}
Here we turned back to the original bi-orthogonal basis in the Hilbert space in order to emphasize the physical meaning of this ground state - the stationary probability distribution given by $P_0 \sim e^{-2W}$. We only consider physically meaningful superpotentials, $W$. In particular, $W$ must be bounded from below. As is seen, $\hat H^\text{cnv}P_0=0$, with the conventional Fokker-Plank operator from Eq.(\ref{CnvFPHam}), or, equivalently, $\hat H|0\rangle\rangle =0$. Unconditional existence of this ground state suggests that $\mathcal Q$-symmetry is never broken for Langevin SDE's with physically meaningful $W$'s.

The above discussion suggests that a model must have a non-potential part in its flow for the $\mathcal Q$-symmetry to be spontaneously broken. In physical terms, to break $\mathcal Q$-symmetry of a Langevin SDE one has to externally "drive" the system through its phase space by a non-potential flow. This driving is actually a well-documented condition for SOC dynamics. \cite{SOC}

It is possible that there exist other classes of models with unbroken $\mathcal Q$-symmetry even in the presence of noise. One of the possible candidates is the conservative models. If it turns out to be true, one would say that a nearly conservative model must possess dissipative part in its drift term for $\mathcal Q$-symmetry to be broken. This is actually one of the conditions for the observation of chaotic behavior in nearly conservative models. Therefore, it is reasonable to expect that in some subclass of stochastic conservative models $\mathcal Q$-symmetry is also unbroken.

\subsection{Quantum excitations vs. stochastic fluctuations}
\label{Explanation}
Here we would like to address the following subtle point. It is known that W-TFT's with unbroken $\mathcal{Q}$-symmetry do not possess physical excitations (see, \emph{e.g.}, Refs. [\onlinecite{Labastida,Anselmi}] as well as Sec.\ref{QsymmetryStates}). This may sound suspicious: how can a theory with no quantum excitations represent a stochastically fluctuating system such as a Langevin SDE? To get around this controversy, in Ref.[\onlinecite{Proposition}] it was wrongfully conjectured that the PSW stochastic quantization is applicable only to stochastic systems with slow colored noises that do not provide high enough frequencies for the system to fluctuate. This is not true - PSW method is applicable to all dynamical models.

The resolution of the seeming controversy is this. Consider again Langevin SDEs as an example. They have only one physical ground state of trivial ghost content - the stationary probability distribution from Eq.(\ref{StationaryDistribution}). This, however, does not mean that there are no stochastic fluctuations. The very necessity to use the probability distribution for the description of the model suggests that the bosonic variables fluctuate. These fluctuations, however, are in the sense of the zero quantum fluctuations within a ground state.

Furthermore, W-TFTs with unbroken $\mathcal Q$-symmetry provide perfectly defined correlators that represent the stochastic fluctuations in the bosonic variables. For example, the one-loop propagator for the fluctuating bosonic fields near an isolated stable critical point, $\varphi_0$, of a (0+1) theory (not necessarily Langevin) is $\langle\langle \varphi_0| \delta\varphi^i(t)\delta\varphi^j(0) | \varphi_0 \rangle\rangle=(\hat D^\dagger\hat D)^{-1}$, where $\hat D =\delta^i_j\partial_t + \partial_{\varphi^j}\hat A^i(\varphi_0)$ and $|\varphi_0\rangle\rangle$ is the one-loop ground state at $\varphi_0$. These correlators are not BPS observables and thus belong to the general class of observables studied in Refs.[\onlinecite{Frenkel}]. The above propagator is exactly the one-loop propagator that would follow from the more wide-spread MSR approximation to stochastic quantization [higher-order perturbative corrections will be different, however, due to the neglect of the contributions from the virtual ghosts in MSR].

\section{Spontaneously broken $\mathcal Q$-symmetry}
\label{SpontanBroken}
In thinking about models with spontaneous $\mathcal Q$-symmetry breaking, it is an important question on which level the symmetry is broken. For any global continuous symmetry, there are three such levels - the mean-filed level, the perturbative level also known as quantum anomaly, and by the dynamical condensation of instantons and anti-instantons. The last possibility is particularly important for supersymmetries such as $\mathcal Q$-symmetry. \cite{DynamicalBreakingOfSUSY} The reason for this is that certain class of supersymmetries, including $\mathcal Q$-symmetry, can not be broken on the perturbative level. This is related to the concepts of the cancelation of bosonic and fermionic determinants, of the localization of path-intergals to classical solutions in W-TFT's (see, \emph{e.g.}, Ref. [\onlinecite{Labastida}]), and to non-renormalization theorems.

This leaves out only two possibilities: $\mathcal Q$-symmetry can be broken either on the mean-field level (which in our case corresponds to the deterministic limit) or by the dynamical condensation of instantons and anti-instantons. Below we argue that these two situations represent respectively chaotic behavior (Sec. \ref{Chaos}) and Intermittent/SOC dynamics (Sec. \ref{SOC}).

Dynamical (anti-)instanton-mediated breaking of $\mathcal Q$-symmetry must not exist in the deterministic limit because of the disappearance of anti-instantons. Nevertheless, in non-equilibrium situations such as quenches instantons are not required to be matched by anti-instantons and consequently instanton(s) alone can break $\mathcal Q$-symmetry even in the deterministic limit.\cite{Frenkel} This possibility is briefly discussed in Sec.\ref{NonEquilibrium}. Let us first, however, address some general aspects of $\mathcal Q$-symmetry broken dynamics.

\subsection{Qualitative difference of $\mathcal Q$-broken dynamics}
\label{QualitativeDifference}

Models with unbroken and spontaneously broken $\mathcal Q$-symmetry must have qualitatively very different dynamics. This difference has a few separate points. Allow us address some of them here.

First, spontaneously broken $\mathcal Q$-symmetry results in the liberation of quantum dynamics in the following sense. If one randomly chooses a wavefunction, $|\Psi\rangle\rangle$, and propagates it in time long enough, he will have:
\begin{eqnarray}\label{evolutionofwavefunction}
|\Psi\rangle\rangle(t) \stackrel{t\to\infty}{\longrightarrow} \sum_{|n\rangle\rangle\in\mathcal{H}^p} e^{-iE_nt}|n\rangle\rangle\langle\langle n| \Psi\rangle\rangle.
\end{eqnarray}
Now it is seen that in case of unbroken $\mathcal Q$-symmetry, when the only physical states are the ground states with $E_n=0$ (see Fig.\ref{Figure2}b), the quantum evolution in Eq.(\ref{evolutionofwavefunction}) stops. In contrary, if $\mathcal Q$-symmetry is spontaneously broken, the time evolution will never stop.

If one would want to construct an effective field theory out of a $\mathcal Q$-broken model, he could as well try to write down a theory that covers the physical states only. Such theory would necessary be Hermitian. In other words, on physical states the Fokker-Plank equation becomes a Sch\"odinger equation without Wick rotation, \emph{i.e.}, preserving the original meaning of time. This line of thinking can be useful for establishing a long-suspected connection between chaotic and quantum dynamics. (see, \emph{e.g.}, Ref. [\onlinecite{ChaosAndQM}]) W-TFT's may provide a firm basis for such connection.

Second, Goldstone theorem ensures that if a global continuous symmetry is spontaneously broken then there must exist a gapless particle. In case of $\mathcal Q$-symmetry, the gapless particles are ghosts called Goldstinos. Goldstinos provide the system with long-range spatiotemporal correlations in some observables related to them. These long-range correlations are known under various names such as sensitivity to initial conditions, self-similarity, algebraic statistics of avalanches in SOC dynamics, 1/f noise,  non-Markovian memory etc. Such long-range correlations must not exist in models with unbroken $\mathcal Q$-symmetry that are thus could be called Markovian dynamics.

Third, the time-reversal symmetry must also be broken in $\mathcal Q$-broken phases as was discussed in the end of Sec.\ref{QsymmetryStates}. This must be related to the concept of irreversibility of, \emph{e.g.}, chaotic dynamics.
\begin{figure}[t] \includegraphics[width=5.0cm,height=2.6cm]{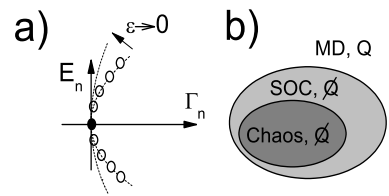}
\caption{\label{Figure4}
{\bf (a)} The spectrum of a toy model considered in Sec.\ref{RefinedDefinitionofChaos}. In the deterministic limit, $\epsilon\to0$, the Fokker-Plank eigenvalues tend to lay on the imaginary axis, as though $\mathcal Q$-symmetry is broken. This situation must also occur in other conservative models. As is explained in the text, this situation must not be viewed as the mean-field level $\mathcal Q$-symmetry breaking. {\bf (b)} The generic phase diagram of a stochastic dynamical system. It consists of three phases. The first is the chaotic phase, where the topological symmetry is broken on the mean-field level. The second phase is intermittent chaos or self-organized criticality, where the topological symmetry is dynamically broken by the condensation of instantons and antiinstantons. SOC is not a "critical state" but rather is a full-dimensional phase. However, in the deterministic limit when antiinstantons disappear, the SOC phase collapses into the critical  "edge of chaos". Chaotic and SOC phases must possess spatiotemporal self-similarity due to the existence of gapless goldstinos. The temporal aspect of the self-similarity represents dynamical non-Markovian long-term memory. By the same token, the phase with unbroken topological symmetry can be called Markovian dynamics (MD).}
\end{figure}

\subsection{Chaotic dynamics}
\label{Chaos}

As it follows from the previous subsection, models with spontaneously broken $\mathcal Q$-symmetry must have qualitatively very different behavior. In deterministic limit, there is only one class of dynamics general enough to be associated with the mean-field level spontaneous $\mathcal Q$-symmetry breaking on the W-TFT side. This class is the deterministic chaos.

In the deterministic limit, the effects of quantum-mechanical tunneling (anti-instantons) between perturbative ground states can be neglected. Perturbative ground states discussed in Sec.\ref{VacuaDetermin} become a good approximation for unbroken $\mathcal Q$-symmetry cases. Then, the rhs of Eq.(\ref{evolutionofwavefunction}) for a randomly chosen initial probability distribution (a wave-function of the maximal degree, $|\Psi\rangle\rangle$) is the sum over all the stable attracting invariant manifolds, while the overlapping coefficients, $\langle\langle n| \Psi\rangle\rangle$, are the integrals of $|\Psi\rangle\rangle$ over the corresponding attracting basins. The absence of quantum evolution means that the deterministic model with unbroken $\mathcal Q$-symmetry will always find its attractor and stay there forever. In contrary, when $\mathcal Q$-symmetry is broken and the evolution nether stops, it must mean that a model is not capable of finding its topologically well-defined attractor - a feature pertinent to chaotic behavior.

Yet another way to see that mean-field $\mathcal Q$-symmetry breaking must be associated with chaos it through considering models with fractal strange attractors. In such cases, one will certainly find it difficult to represent such an invariant "manifold" (formally a fractal is not a topological manifold) as an intersection of forms, that the ground states are supposed to be in the unbroken $\mathcal Q$-symmetry models. The first problem here is the differentiability issue. The second problem is more fundamental. The point is that states can only be forms of integer degree. Such forms can only represent integer dimensional manifolds but not fractals.

Looking at deterministic chaotic models as at the mean-field level $\mathcal Q$-symmetry breaking suggests why they must exhibit long-range correlations in observables related to ghosts. These correlations is the W-TFT way to explain the high sensitivity to initial conditions in chaotic models. In dynamical systems theory, this sensitivity is often considered as one of the necessary conditions for the system to be chaotic. On the W-TFT side, the sensitivity is the consequence of the Goldstone theorem. Furthermore, time-reversal symmetry of deterministic chaos must also be broken. This is most probably the W-TFT way of encoding the temporal irreversibility of chaos. \cite{Irreversibility}

The above reasons are only indications on the equivalence between deterministic chaos and mean-field level spontaneous $\mathcal Q$-symmetry breaking. A rigorous prove of this equivalence has obstacles. One of them is the absence of the universal definition of deterministic chaos. From this perspective, one can as well think that W-TFT's approach provides its own version of this definition.

If $\mathcal Q$-symmetry is broken in the deterministic limit, \emph{i.e.}, on the mean-field level, the addition of noise will not restore it. In fact, as is discussed in the next subsection, the noise can only break $\mathcal Q$-symmetry if it was not broken on the mean-field level. Stochastic models with $\mathcal Q$-symmetry broken on the mean-field level can be called stochastic chaos.

\subsubsection{Refined definition of deterministic chaos}
\label{RefinedDefinitionofChaos}
So far, the W-TFT definition of chaos is a model with $\mathcal Q$-symmetry spontaneously broken on the mean-field level. This definition needs further clarification. The reason for this is seen from the following example.

Consider a dynamical model on a circle, $\mathbb{S}^1$, of circumference, $L$, with constant flow, $A$. The operators defining the Fokker-Plank Hamiltonian through Eq.(\ref{FPHamiltonian}) are:
\begin{eqnarray}
\hat Q = \chi\hat \partial_\varphi, \hat{\bar Q} = \partial_\chi( - \epsilon\partial_\varphi - 2 A).
\end{eqnarray}
The eigenstates and the spectrum are known exactly:
\begin{eqnarray}
|\Psi\rangle\rangle^{(0)}_{k} = \chi e^{ik\varphi}, |\Psi\rangle\rangle^{(1)}_{k} = e^{ik\varphi},\\
\mathcal{E}^{(0)}_k = \mathcal{E}^{(1)}_k = \epsilon k^2/2 - 2 i k A,
\end{eqnarray}
where $\varphi\in[0,L]$ and the momentum is $k L/(2\pi)\in \mathbb{Z}$. As is seen in Fig.\ref{Figure4}a, the spectrum looks like that of broken $\mathcal Q$-symmetry in the strict deterministic limit, $\epsilon= 0$. On the other hand, for any small but finite noise, $0<\epsilon \ll 1$, the spectrum is that of unbroken $\mathcal Q$-symmetry. Such situation must appear in all conservative models.

To get around this seeming controversy, we have to refine the definition of the mean-field level $\mathcal Q$-symmetry breaking and thus that of chaos. Chaotic system is such that its $\mathcal Q$-symmetry is spontaneously broken in the limit of infinitely weak but finite noises. This weak but finite noise is related to the concept of sensitivity to initial condition on the dynamical systems side. To see if a deterministic model is sensitive to initial conditions one must consider infinitely close but still different initial points in the phase space. This can be interpreted as the addition of infinitely weak but still finite uncertainty (noise).

\subsection{Intermittency/SOC}
\label{SOC}

In stochastic models, there is yet another mechanism for spontaneous $\mathcal Q$-symmetry breaking. This mechanism is the dynamical condensation of instantons and anti-instantons. \cite{DynamicalBreakingOfSUSY,Affleck,InstantonBRSTBreaking} We have already encountered instantons in Sec.\ref{VacuaDetermin} in the discussion of the ground states in the deterministic limit. Let us, however, briefly recollect again on what these tunneling processes are.
Instatnons correspond to classical solutions of SDE (solution of a corresponding DDE) that start on one invariant manifold and end at another. Thus, instantons always lead from "less stable" invariant manifolds to more stable ones. Anti-instantons, in turn, are time-reversed instantons. They lead from more stable invariant manifolds to less stable ones. Anti-instantons are essentially the motion against the flow, which is only possible in the presence of noise. Matrix elements of anti-instantons always come with exponentially small factors that disappear in the deterministic limit. A hydrodynamical example of an instanton/antiinstanton is the processes of annihilation/creation of (pairs of) vortices or vortex lines.

(Anti-)instantons intrinsically break $\mathcal Q$-symmetry as, in particular, is seen in Eq.(\ref{Morse-Witten}), from which it follows that the matrix element of the expectation value of a $\mathcal Q$-symmetry operator on the $(ba)$-instanton (see Fig.\ref{Figure3}b) is non-zero: $\langle\langle a| \hat Q| b\rangle\rangle = 1$.

In equilibrium situations, condensation of instantons must be accompanied by the condensation of antiinstantons. Indeed, for the instantonic processes of annihilation of vortices to be happening forever they must be complemented by the antiinstantonic processes of creation of vortices. In other words, separately each (anti-)instanton is ultimately non-equilibrium processes leading from one state to another. Therefore, in equilibrium situations, only configurations of "equal number" of instantons and anti-instantons that lead from a state to itself exist. Thus, in equilibrium, (anti-)instanton-mediated dynamical $\mathcal Q$-symmetry breaking can only happen in stochastic but not deterministic models.

The dynamics in a phase with the (anti-)instanton-mediated $\mathcal Q$-symmetry breaking must look as an infinite series of jumps (\emph{e.g.}, avalanches of sandpile models) between different invariant manifolds. If the invariant manifolds are points in the phase space (solitonic configurations, patterns etc.), we can talk about the dynamics as of a sequence of sudden changes in solitonic configuration. This is nothing else but the physical picture of SOC. \cite{SOC}

In a more general case, these jumps happen not between points but between more intricate invariant manifolds (\emph{e.g.}, limit cycles). In this case, the term SOC must be substituted by a more general concept of Intermittency (see. e.g., Ref.\cite{Intermittency}) - in dynamical systems theory a system is called intermittent if it is subject to infrequent variations of large amplitude. These variations ((anti-)instantons) separate periods of different (quasi-)periodic behaviors. Roughly speaking, the difference between Intermittency and SOC is that the jumps happen between limits cycles or more complicated higher-dimensional stable sets for the former and between stable points for the later. Thus, SOC is in a sense a subclass of intermittent behavior. We do not see, however, more fundamental difference between SOC and Intermittency and thus sometimes use these terms interchangeably.

The conclusion we just arrived is that (anti-)instanton-mediated $\mathcal Q$-symmetry broken phase must be associated with Intermitency/SOC. \cite{Proposition} On the phase diagram, intermittent/SOC phase must occupy a region between chaotic and Markovian phases. Importantly, Intermittency/SOC is not a critical state. It is a rightful full-dimensional phase, just as chaos and Markovian systems. This explain why by moderate variation of parameters of the model one can not destroy intermittent/SOC behavior. In the literature of SOC, this fact is sometimes attributed to the mysterious tendency of SOC systems to "evolve" to a critical "edge of chaos". W-TFT picture of Intermittency/SOC resolves this issue.

Only in the deterministic limit, anti-instantons vanish and intermittent/SOC phase collapses into the "edge of chaos" (see, \emph{e.g.}, Ref.[\onlinecite{EdgeOfChaos}]). The above discussion is summarized in Fig.(\ref{Figure4}b) where chaotic, intermittent/SOC, and Markovian phases are schematically presented.

The proposed phase diagram complies with previous studies. One of the examples is the phase diagram of water moving between two rotating cylinders. \cite{PhaseDiagram} The Couette (laminar) flow phase can be identified with Markovian phase, the featureless turbulence with chaos, while the intermediate region, where the dynamics is dominated by creation/annihilation of solitons in the form of vortices and vortex lines, must be identified with intermittency/SOC. Yet another example is neuroscience, where the three phases in Fig.(\ref{Figure4}b) are often called respectively supercritical, critical, and subcritical phases. \cite{Levina}

\subsection{Non-equilibrium dynamics}
\label{NonEquilibrium}
Let us also briefly address non-equilibrium situations. In these situations, the temporal boundary conditions are not periodic and the system is allowed to end up in a state different from the initial state. Physically, this corresponds to quenches for instance. Quench dynamics can be assumed deterministic and it can be looked upon as a composite instanton leading from one of the metastable configurations to one of the stable configurations. Barkhausen-like effects (see, \emph{e.g.}, Ref.[\onlinecite{Barkhausen}] and Refs. therein) such as crumpling paper (see, \emph{e.g.}, Ref.[\onlinecite{CrumplingPaper}] and Refs. therein) can effectively be viewed as "slow" quenches where the time evolution is due to varying external parameter. Gradual variation of the external parameter (\emph{e.g.}, magnetic field in Barkhausen effect) makes previously stable configurations unstable ones thus initiating instantons, that look like a sequence of sudden jumps. As follows from the next paragraph, these models must also exhibit "power-laws" which is the reason why they also sometimes attributed to the SOC family.

Mathematical aspects of non-equilibrium deterministic dynamics in potential flows (Langevin DDE's) were studied, \emph{e.g.}, in Ref.[\onlinecite{Frenkel}]. There, it was found that the corresponding theory is a log-conformal theory. Roughly, one can attribute the long-range log-conformal correlations to the $\mathcal Q$-symmetry spontaneously broken by a given composite instanton. Interestingly, a quench must not necessary be across some-other-symmetry breaking phase transition in order to exhibit long-range correlations, \emph{e.g.}, Barkhausen-like dynamics. It is also interesting that for quenches one can straightforwardly use BPS observables that may provide certain topological invariants in topologically non-trivial theories. It is tempting to believe that so-calculated topological invariants may have something to do with the probabilities of the system to end up in one or the other competing stable configurations.

\section{Conclusion}
\label{Conclusion}

In this paper, it is shown that the most general stochastic quantization procedure applied to any stochastic or deterministic continuous-time (partial) differential equations leads to a Witten-type topological field theory - a theory with global topological supersymmetry. This topological symmetry must be perturbatively stable and consequently can be spontaneously broken only either on the mean-field level by, say, fractal invariant sets, or dynamically by the condensation of instantons and anti-instantons. According to this, we propose a generic phase diagram given in Fig.(\ref{Figure1}a). The mean-field level spontaneous $\mathcal Q$-breaking must be associated with chaotic behavior. The instanton-anti-instanton mediated breaking of topological symmetry corresponds to intermittency/SOC - a full dimensional phase surrounding chaos. Its full-dimensionality explains why algebraic correlations of SOC dynamics are robust against moderate variation of the parameters of a system. Full-dimensional intermittency/SOC phase exists, however, only in stochastic models. In deterministic models, when anti-instantons disappear, intermittency/SOC phase degenerates into the critical "edge of chaos" between deterministic chaos and deterministic Markovian dynamics.

According to Goldstone theorem, both phases with spontaneously broken topological symmetry must exhibit long-range correlations. These correlations are related to such well-established concepts as self-similarity, sensitivity to initial conditions and/or non-Markovian scale-free memory, power-law statistics of avalanches, algebraic power-spectrum in turbulence etc. Accordingly, dynamical systems with unbroken topological symmetry can be called Markovian in that sense that they do not exhibit long-range dynamical memory so one can always look at the system at such temporal scale that the dynamics will look effectively Markovian. \cite{vanKampen}

The phase diagram and the interpretation of the physical essence of the three phases is the only result of this paper. There is certainly much more in the W-TFT of dynamical systems to be understood in the future. In particular, in this paper we never used one of the most fascinating possibilities that W-TFT's offer - the possibility to calculate certain topological invariants as expectation values of BPS observables on instantons. One of the candidate ways to exploit these observables (probably in combination with their anti-instantonic counterparts) is in the form of order parameters for $\mathcal Q$-broken phases. In fact, a rigorous way for construction of effective field theories is one of the important advancements that W-TFT may lead to. W-TFT of dynamical systems may also hold the key to some novel
  topological aspects of probability theory.

\acknowledgements
We would like to thank R. Rohwer, C. Connolly, J. Wang, H.-H. Shieh, R. Kozma, W. Freeman, A. Stieg, J. Gimzewski, B. Jalali, R. Schwartz, and K. L. Wang for stimulating discussions. We would also like to thank T. Hylton for bringing the subject to our attention. The work was supported by Defense Advanced Research Projects Agency, Defense Sciences Office, Program: Physical Intelligence, contract HR0011-01-1-0008.

\end{document}